\documentclass[aps,prb,twocolumn, superscriptaddress]{revtex4-2}
\usepackage{float}
\usepackage{hyperref}
\usepackage{graphicx}
\usepackage[utf8]{inputenc}
\DeclareGraphicsExtensions{.png,.jpg,.eps,.pdf}
\usepackage{amsmath}
\usepackage{comment}
\usepackage{bm}
\usepackage{braket}
\usepackage{dsfont}
\usepackage{amsfonts}
\usepackage{xcolor}
\usepackage{bbold}
\usepackage{physics}
\usepackage[normalem]{ulem}
\usepackage{tikz}
\usetikzlibrary{quantikz,trees,arrows,positioning,automata,shadows,fit,shapes,cd}

\begin{document}



\title{From Heisenberg to Hubbard: An initial state for the \\ shallow quantum simulation of correlated electrons}

\author{Bruno Murta}
\email{bpmurta@gmail.com}
\affiliation{Departamento de F\'{i}sica, Universidade do Minho, Campus de Gualtar, 4710-057 Braga, Portugal }
\affiliation{International Iberian Nanotechnology Laboratory (INL), Av. Mestre Jos\'{e} Veiga, 4715-330 Braga, Portugal}

\author{J. Fern\'{a}ndez-Rossier}
\altaffiliation{On permanent leave from Departamento de F\'{i}sica Aplicada, Universidad de Alicante, 03690 San Vicente del Raspeig, Spain}
\affiliation{International Iberian Nanotechnology Laboratory (INL), Av. Mestre Jos\'{e} Veiga, 4715-330 Braga, Portugal}

\date{\today}


\begin{abstract}

The widespread use of the noninteracting ground state as the initial state for the digital quantum simulation of the Fermi-Hubbard model is largely due to the scarcity of alternative easy-to-prepare approximations to the exact ground state in the literature. Exploiting the fact that the spin-$\frac{1}{2}$ Heisenberg model is the effective low-energy theory of the Fermi-Hubbard model at half-filling in the strongly interacting limit, here we propose a three-step deterministic quantum routine to prepare an educated guess of the ground state of the Fermi-Hubbard model through a shallow circuit suitable for near-term quantum hardware. First, the ground state of the Heisenberg model is initialized via a hybrid variational method using an ansatz that explores only the correct symmetry subspace. Second, a general method is devised to convert a multi-spin-$\frac{1}{2}$ wave function into its fermionic version. Third, taking inspiration from the Baeriswyl ansatz, a constant-depth single-parameter layer that adds doublon-holon pairs is applied to this fermionic state. Numerical simulations on chains and ladders with up to $12$ sites confirm the improvement over the noninteracting ground state of the overlap with the exact ground state for the intermediate values of the interaction strength at which quantum simulation is bound to be most relevant. 

\end{abstract}

\maketitle

Digital quantum simulation~\cite{McArdle19, Cao19} is expected to become a leading method to study correlated electrons~\cite{Auerbach94}. By exploiting the principle of superposition and the natural encoding of entanglement, quantum computers can represent the full wave function of quantum many-body systems in a scalable way, which may allow to probe properties that defy state-of-the-art numerical methods on conventional hardware~\cite{Foulkes01, Schollwock05}. A problem that offers the prospect of achieving such quantum advantage~\cite{Cade20, Cai20} even with noisy intermediate-scale quantum (NISQ) processors~\cite{Preskill18} is the determination of the phase diagram of the Fermi-Hubbard model~\cite{Gutzwiller63, Hubbard63, Kanamori63} by preparing the exact ground state of the second-quantized Hamiltonian
\begin{equation}
    \hat{\mathcal{H}} = -t \sum_{i, \tau} \sum_{\sigma = \uparrow, \downarrow} ( \hat{c}^{\dagger}_{i, \sigma} \hat{c}_{i+\tau, \sigma} + \textrm{H.c.} ) + U \sum_{i} \hat{n}_{i, \uparrow} \hat{n}_{i, \downarrow},
    \label{eq:Fermi_Hubbard_Hamiltonian}
\end{equation}
where the sum over $\tau$ includes the nearest neighbors of site $i$, and $t > 0$ and $U > 0$ define the interaction strength $\frac{U}{t}$. The most challenging and relevant regime~\cite{Arovas22} of the Fermi-Hubbard model occurs when the two competing energy scales are comparable --- i.e., the Hubbard parameter $U$ is of the order of the bandwidth $W$ of the underlying tight-binding model (e.g., $W = 4t$ in one dimension, $W = 8t$ for the square lattice).

A key requirement for any ground state preparation method is an initial state with non-negligible overlap with the target state. In the case of the Fermi-Hubbard model, the standard choice is the noninteracting ground state~\cite{Cade20, Cai20}, but its vanishing fidelity relative to the exact ground state~\cite{Murta21} for the intermediate range $U \sim W$ calls for a more educated guess. Mean-field states~\cite{Wecker15} face the same issue, with the additional drawback of often breaking symmetries of the Hamiltonian. The Gutzwiller wave function~\cite{Gutzwiller63} does produce a substantially greater overlap with the exact ground state at intermediate and large $\frac{U}{t}$, but the NISQ-friendly schemes proposed to initialize it~\cite{Murta21, Seki22} require, on average, a number of repetitions to succeed that becomes prohibitively large for a lattice of sufficiently great size due to their probabilistic nature. 

In this Letter we introduce a deterministic quantum routine that is suitable for NISQ hardware to prepare a better approximation than the noninteracting ground state of the exact ground state of the Fermi-Hubbard model at half-filling with intermediate or large $\frac{U}{t}$. This scheme makes use of the fact that, in the strongly interacting limit $\frac{U}{t} \to \infty$, the charge degrees of freedom are frozen and the Fermi-Hubbard model is reduced~\cite{Auerbach94} to the antiferromagnetic spin-$\frac{1}{2}$ Heisenberg model~\cite{Heisenberg28}
\begin{equation}
    \hat{\mathcal{H}} = J \sum_{i, \tau} \left( \hat{S}^{x}_{i} \hat{S}^{x}_{i+\tau} + \hat{S}^{y}_{i} \hat{S}^{y}_{i+\tau} + \hat{S}^{z}_{i} \hat{S}^{z}_{i+\tau} \right),
\label{eq:Heisenberg_Hamiltonian}
\end{equation}
with $J = \frac{4t^2}{U}$. This result is valid for any lattice at half-filling and may be extended to hopping terms beyond nearest neighbors. Although determining the ground state of the Heisenberg model is generally nontrivial, we can benefit from the smaller size of the Hilbert space relative to the full-blown fermionic model to mitigate some of the most cumbersome issues faced in quantum simulation that arise from the exponential wall problem~\cite{Kohn99}, namely the orthogonality catastrophe~\cite{Anderson67} and the barren plateaus~\cite{McClean18} in hybrid variational methods~\cite{Cerezo21}. 

\begin{figure}[t]
\includegraphics[width=\linewidth]{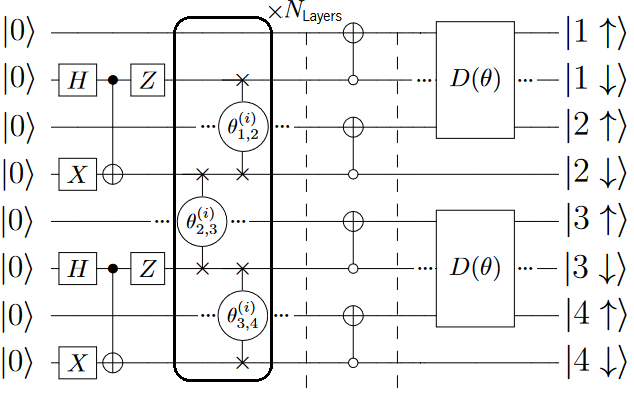}
\caption{Quantum circuit that prepares improved fermionic version of ground state of spin-$\frac{1}{2}$ Heisenberg model~\cite{Heisenberg28} on a lattice with $N=4$ sites, which approximates the ground state of the Fermi-Hubbard model~\cite{Gutzwiller63, Hubbard63, Kanamori63}. The first part (left of first dashed line) prepares the ground state of the Heisenberg model via VQE~\cite{Peruzzo14} using a $N_{\textrm{Layers}}$-layer ansatz that starts from a product state of valence bonds and explores the $S_{\textrm{Total}} = 0$ subspace~\cite{Seki20}. The two-qubit building block of the ansatz is defined in Eq. (\ref{eq:eSWAP}). The second part converts this wave function into its $2N$-qubit fermionic version, assuming the Jordan-Wigner transformation~\cite{JordanWigner28} is used. The third part (right of second dashed line) introduces pairs of empty and doubly-occupied sites to this fermionic state. The two-qubit operation $D(\theta)$ is defined in Eq. (\ref{eq:D_theta_def}). The ellipses before and after two-qubit operations indicate the middle qubit is idle.}
\label{fig1}
\end{figure}

The quantum scheme herein put forth comprises three parts that can be identified in the circuit scheme shown in Fig. \ref{fig1}. The first part concerns the preparation of the ground state of the spin-$\frac{1}{2}$ Heisenberg model defined on a $N$-site balanced bipartite lattice via the Variational Quantum Eigensolver (VQE)~\cite{Peruzzo14} with an ansatz that explores only the $S_{\textrm{Total}} = 0$ subspace~\cite{Seki20}. The second part converts this $N$-spin-$\frac{1}{2}$ wave function into the respective fermionic state defined on $2N$ qubits, assuming the Jordan-Wigner transformation~\cite{JordanWigner28} is employed. The third part introduces pairs of empty and doubly-occupied sites in the fermionic version of the Heisenberg ground state. The remainder of this Letter will detail each step, with numerical results of simulations on chains and ladders with up to $12$ sites complementing the explanations.

In the first part, the exact ground state of the antiferromagnetic spin-$\frac{1}{2}$ Heisenberg model on a given lattice with $N$ sites is prepared via VQE. The building block of the ansatz is the $\textrm{eSWAP}(\theta) = e^{-i \frac{\theta}{2} \textrm{SWAP}}$~\cite{Seki20}, corresponding to the following SU(2)-invariant two-qubit operation
\begin{equation}
    \begin{pmatrix}
    e^{-i \frac{\theta}{2}} & 0 & 0 & 0 \\
    0 & \cos \frac{\theta}{2} & -i \sin \frac{\theta}{2} & 0 \\
    0 & -i \sin \frac{\theta}{2} & \cos \frac{\theta}{2} & 0 \\
    0 & 0 & 0 & e^{-i \frac{\theta}{2}}
    \end{pmatrix} \; \equiv 
    \raisebox{-0.42\totalheight}{
    \includegraphics[width=0.1\linewidth]{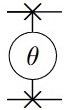}
    }.
\label{eq:eSWAP}
\end{equation}
Starting from a reference state with the expected $S_{\textrm{Total}}$ and $S^{z}_{\textrm{Total}}$ for the exact ground state ensures the manifold spanned by the parameterized state is confined to the subspace defined by these good quantum numbers. In particular, for a balanced bipartite lattice, the Lieb-Mattis theorem~\cite{LiebMattis62} guarantees that the ground state of the antiferromagnetic spin-$\frac{1}{2}$ Heisenberg model has $S_{\textrm{Total}} = S^{z}_{\textrm{Total}} = 0$, so an appropriate reference state motivated by the ease of preparation is a product state of valence bonds $\frac{\ket{\uparrow \downarrow} - \ket{\downarrow \uparrow}}{\sqrt{2}}$ at the odd-even pairs $(1,2), (3,4), ..., (N-1,N)$ of adjacent qubits in a linear configuration. On top of this reference state, $N_{\textrm{layers}}$ layers of the ansatz are applied, each corresponding to the execution of $\frac{N}{2}-1$ eSWAPs at the even-odd pairs of qubits, followed by the implementation of $\frac{N}{2}$ eSWAPs at the odd-even pairs (see Fig. \ref{fig1}). The structure of the ansatz is defined such that even in the most restrictive case of linear qubit connectivity no rerouting of qubits is required to implement it. Although previous works~\cite{BosseMontanaro22, YuZhaoWei23} have motivated this ansatz in the spirit of discretized adiabatic evolution~\cite{Farhi14, Wecker15a}, here we follow its interpretation~\cite{Seki20} as a generator of resonating-valence-bond (RVB) states~\cite{Anderson73, Anderson87}, in which case each eSWAP is assigned its own free parameter, resulting in a total of $N_{\textrm{layers}}(N-1)$ parameters. Henceforth, this ansatz will be referred to as \textit{RVB-inspired ansatz}.

\begin{figure*}[t]
\centering
\includegraphics[width=\linewidth]{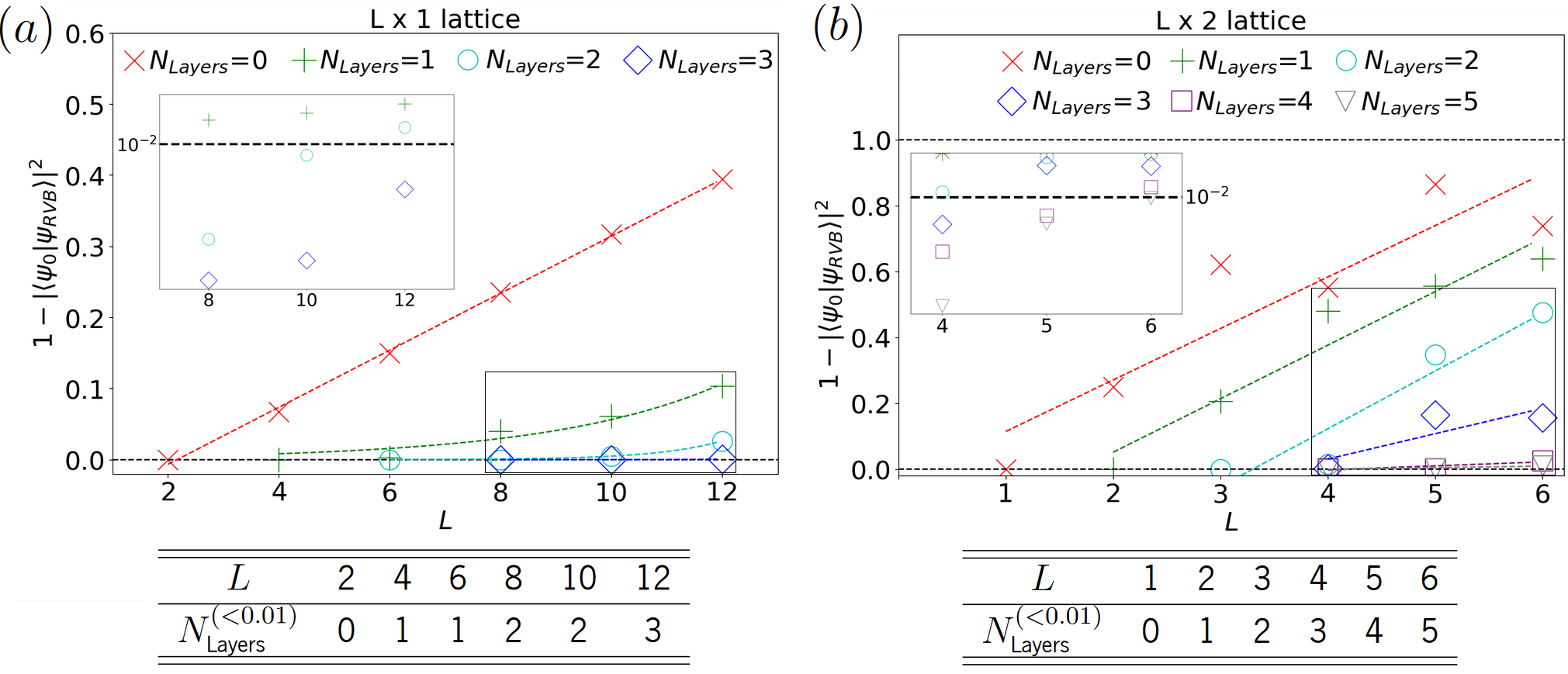}
\caption{Infidelity between exact ground state and RVB-inspired ansatz \cite{Seki20} obtained via VQE \cite{Peruzzo14} for antiferromagnetic spin-$\frac{1}{2}$ Heisenberg model \cite{Heisenberg28} in chains (a) and ladders (b) with up to $12$ sites and open boundary conditions. The inset plots are a zoomed-in view (with the vertical axis in logarithmic scale) of the highlighted rectangular section in the bottom-right corner of the main plot. The tables below each graph present the minimum number of layers of the RVB-inspired ansatz to achieve an infidelity below $0.01$ with respect to the exact ground state. The dashed lines in the main plots are just a guide to the eye. As expected, as the number of lattice sites increases, more ansatz layers are required to achieve a given infidelity target. $N_{\textrm{Layers}} = 0$ corresponds to the product state of valence bonds that is the input state of the RVB-inspired ansatz (see Fig. \ref{fig1}).}
\label{fig2}
\end{figure*}

Since the first step is the only potential bottleneck of this quantum scheme, \textit{in silico} noiseless simulations were performed to assess the scalability of the preparation of the ground state of the spin-$\frac{1}{2}$ Heisenberg model via VQE. The basis gate decompositions and the details of the optimization of the parameters of the ansatz can be found in the Supp. Mat.~\cite{SuppMatRef}. Two types of lattice geometries were considered: $L \times 1$ (i.e., chains) and $L \times 2$ (i.e., ladders). The total number of lattice sites --- $N = L$ and $N = 2L$, respectively --- was at most $12$. Open boundary conditions were considered in all cases, but periodic boundary conditions could have been adopted just as well. The total number of sites was always chosen to be even to ensure that the ground state is a singlet. The true ground state was determined via exact diagonalization \cite{QuSpin}, which allowed to calculate the infidelity, $1 - |\langle \psi_0 | \psi_{\textrm{RVB}} \rangle|^2$, between the exact ground state, $\ket{\psi_0}$, and the optimized RVB-inspired ansatz, $\ket{\psi_{\textrm{RVB}}}$, for a given $N_{\textrm{Layers}}$. The results are shown in Fig. \ref{fig2}. As expected, for a fixed $N_{\textrm{Layers}}$, the infidelity relative to the exact ground state increases as the lattice size grows. Conversely, increasing $N_{\textrm{Layers}}$ for a given lattice produces a closer approximation to the true ground state. Importantly, as shown in the tables below the respective graphs, the minimum number of layers required to achieve an infidelity of at most $0.01$ increases linearly with the number of lattice sites, and the prefactor is small. For example, the $3$-layer RVB-inspired ansatz that approximates the exact ground state of the Heisenberg model on a $12$-site chain with fidelity $0.9993$ takes only $19$ CNOTs of depth, which is less than the $22$ CNOTs of depth required to prepare the noninteracting ground state of the corresponding Fermi-Hubbard model~\cite{Kivlichan18, Jiang18}. As for the $6 \times 2$ ladder, the $5$ layers of the RVB-inspired ansatz that produce a fidelity of $0.991$ take $31$ CNOTs of depth, which is not significantly above the $22$ CNOTs needed to prepare the noninteracting ground state of the Fermi-Hubbard model~\cite{Kivlichan18, Jiang18}. Hence, assuming the observed trend continues for larger lattices, the resulting circuits should be shallow enough for NISQ hardware.

We now proceed to the second part of the quantum scheme, where the $N$-spin-$\frac{1}{2}$ ground state of the Heisenberg model is converted into a $2N$-qubit fermionic state that is the exact ground state of the Fermi-Hubbard model on the same lattice at half-filling in the $\frac{U}{t} \to \infty$ limit. This conversion is valid for any multi-spin-$\frac{1}{2}$ state.

Let us first consider a generic single-spin-$\frac{1}{2}$ state $\ket{\psi^{\textrm{spin}}} = a \ket{\uparrow} + b \ket{\downarrow} \equiv a \ket{0} + b \ket{1}$. The corresponding single-site fermionic state is $\ket{\psi^{\textrm{fermion}}} = a \ket{10} + b \ket{01}$, where the Jordan-Wigner transformation~\cite{JordanWigner28} is implicit. If we add an ancillary qubit in $\ket{0}$ to $\ket{\psi^{\textrm{spin}}}$, as in $\ket{0} \otimes \ket{\psi^{\textrm{spin}}} = a \ket{00} + b \ket{01}$, the two-qubit operation that must be applied to transform $\ket{0} \otimes \ket{\psi^{\textrm{spin}}}$ into $\ket{\psi^{\textrm{fermion}}}$ should map $\ket{00}$ to $\ket{10}$ whilst leaving $\ket{01}$ unchanged. The action of this unitary operation on the remaining two basis states is immaterial. Hence, a valid choice is
\begin{equation}
    \begin{pmatrix}
    0 & 0 & 1 & 0 \\
    0 & 1 & 0 & 0 \\
    1 & 0 & 0 & 0 \\
    0 & 0 & 0 & 1
    \end{pmatrix} \quad \equiv 
    \raisebox{-0.42\totalheight}{
    \begin{tikzpicture}
    \node[scale=1.1]{
    \begin{tikzcd}[row sep={0.6cm,between origins}, column sep=0.1cm]
    & \targ{} & \qw \\
    & \octrl{-1} & \qw
    \end{tikzcd}
    };
    \end{tikzpicture}},
\label{eq:spin_to_fermion_qc}
\end{equation}
where the top qubit in the diagram is the most significant and the unfilled circle means the NOT gate is triggered in the target-qubit only when the control-qubit is in $\ket{0}$. 

\begin{figure*}[t]
\centering
\includegraphics[width=\linewidth]{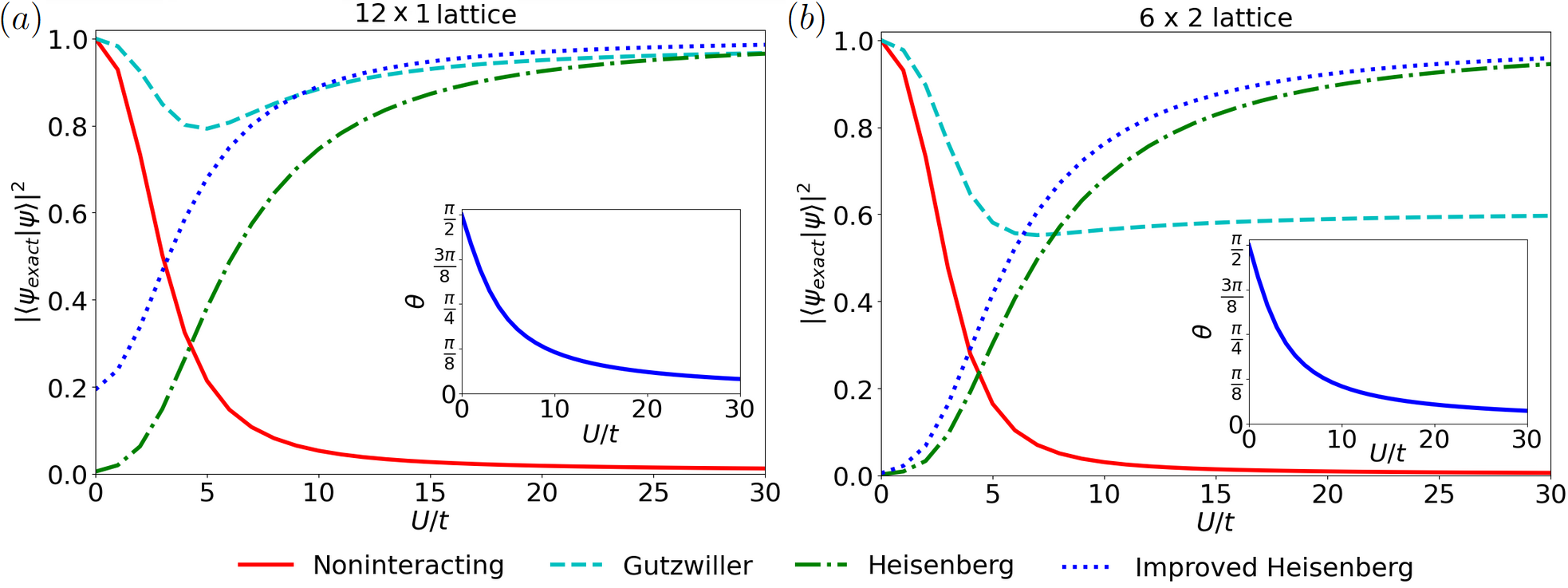}
\caption{Fidelity relative to exact ground state of Fermi-Hubbard model of noninteracting ground state (red solid line), Gutzwiller wave function \cite{Gutzwiller63} (cyan dashed line), fermionic version of Heisenberg ground state (green dashed-dotted line), and its improved version via the addition of doublon-holon pairs with a layer with a single free parameter $\theta$ (blue dotted line) for $12 \times 1$ (a) and $6 \times 2$ (b) lattices. The RVB-inspired ansatz with the minimum number of layers to achieve a fidelity of at least $0.99$ was used to initialize the Heisenberg ground state (see Fig. \ref{fig2}). Open boundary conditions were considered. The insets show the optimized free parameter $\theta$ against the interaction strength $\frac{U}{t}$. Identical $\theta \textrm{ vs. } \frac{U}{t}$ were obtained for smaller lattices. In fact, the values of $\theta$ used for these $12$-site-lattice simulations were obtained from the simulations with $10$-site lattices of the same geometry.}
\label{fig3}
\end{figure*}

The generalization to an arbitrary number of lattice sites is straightforward apart from the extra minus signs due to the anti-commutation relations of fermionic operators. Actually, these fermionic signs only arise when the qubits are ordered by spin instead of site. Indeed, in the latter case, the amplitudes of $\ket{\psi^{\textrm{fermion}}}$ are exactly equal to those of $\ket{\psi^{\textrm{spin}}}$. For example, a basis state of the form $\ket{\uparrow \downarrow \downarrow \uparrow \downarrow \uparrow}$ corresponds explicitly to $\ket{\uparrow}_{1} \otimes \ket{\downarrow}_{2} \otimes \ket{\downarrow}_{3} \otimes \ket{\uparrow}_{4} \otimes \ket{\downarrow}_{5} \otimes \ket{\uparrow}_{6}$, the fermionic version of which is naturally ordered by site as well: $\hat{c}^{\dagger}_{1, \uparrow} \hat{c}^{\dagger}_{2, \downarrow} \hat{c}^{\dagger}_{3, \downarrow} \hat{c}^{\dagger}_{4, \uparrow} \hat{c}^{\dagger}_{5, \downarrow} \hat{c}^{\dagger}_{6, \uparrow} \ket{\Omega}$. As a result, ordering the $2N$ qubits by site, the conversion of $\ket{\psi^{\textrm{spin}}}$ into $\ket{\psi^{\textrm{fermion}}}$ amounts to repeating the application of the two-qubit operation stated in Eq. (\ref{eq:spin_to_fermion_qc}) at every pair of qubits encoding a lattice site (see Fig. \ref{fig1}). If the qubits are ordered by spin instead, returning to the example above, upon commuting the creation operators we obtain $-\hat{c}^{\dagger}_{1, \uparrow} \hat{c}^{\dagger}_{4, \uparrow} \hat{c}^{\dagger}_{6, \uparrow} \hat{c}^{\dagger}_{2, \downarrow} \hat{c}^{\dagger}_{3, \downarrow} \hat{c}^{\dagger}_{5, \downarrow} \ket{\Omega}$, so an extra minus sign must be applied to its amplitude. If $S^{z}_{\textrm{Total}}$ is a good quantum number, these fermionic signs can be accounted for by replacing Eq. (\ref{eq:spin_to_fermion_qc}) at every other site with
\begin{equation}
    \begin{pmatrix}
    0 & 0 & 1 & 0 \\
    0 & 1 & 0 & 0 \\
    -1 & 0 & 0 & 0 \\
    0 & 0 & 0 & 1
    \end{pmatrix} \; \equiv 
    \raisebox{-0.42\totalheight}{
    \begin{tikzpicture}
    \node[scale=1.1]{
    \begin{tikzcd}[row sep={0.6cm,between origins}, column sep=0.1cm]
    & \gate{X} & \gate{H} & \targ{} & \gate{H} & \gate{X} & \targ{} & \qw \\
    & \qw & \qw & \octrl{-1} & \qw & \qw & \octrl{-1} & \qw 
    \end{tikzcd}
    };
    \end{tikzpicture}},
\label{eq:spin_to_fermion_qc_2}
\end{equation}
resulting in a constant-depth overhead as well. In case the input state involves basis states with different $S^{z}_{\textrm{Total}}$, a network of fermionic SWAPs~\cite{Verstraete09} is required to prepare the spin-ordered fermionic state (see Supp. Mat. \cite{SuppMatRef}). 

The third and final part of the quantum routine aims to increase the overlap of the fermionic Heisenberg ground state $\ket{\psi^{(U/t \to \infty)}_{0}}$ with the exact ground state of the Fermi-Hubbard model at finite $\frac{U}{t}$. A layer of two-qubit operations $D(\theta)$ with a single free parameter $\theta$ is applied to $\ket{\psi^{(U/t \to \infty)}_{0}}$. In the spirit of the Baeriswyl wave function~\cite{Baeriswyl00}, this layer promotes the hopping of spin-$\uparrow$ electrons between adjacent odd-even pairs of sites (see Fig. \ref{fig1}) to give rise to doublon-holon pairs:
\begin{equation}
    \begin{aligned}
    & D(\theta) \ket{\uparrow, \uparrow} = \ket{\uparrow, \uparrow}, \; \; \; D(\theta) \ket{\downarrow, \downarrow} = \ket{\downarrow, \downarrow}, \\
    & D(\theta) \ket{\uparrow, \downarrow} = \cos \frac{\theta}{2} \ket{\uparrow, \downarrow} + \sin \frac{\theta}{2} \ket{0, \uparrow \downarrow}, \\
    & D(\theta) \ket{\downarrow, \uparrow} = \cos \frac{\theta}{2} \ket{\downarrow, \uparrow} - \sin \frac{\theta}{2} \ket{\uparrow \downarrow, 0}.
    \end{aligned}
\label{eq:D_theta_def}
\end{equation}
Crucially, the expected improvement of the fidelity with the exact ground state is only observed if $D(\theta)$ is applied to the qubits encoding the $\ket{i, \uparrow}$ and $\ket{i + 1, \uparrow}$ orbitals with $\ket{i, \downarrow}$ between them (see Fig. \ref{fig1}). Expanding $D(\theta)$ in the Pauli basis and applying the Jordan-Wigner transformation~\cite{JordanWigner28} in reverse, we obtain in this case
\begin{equation}
    \begin{aligned}
    & D(\theta) = \frac{1 + \cos \frac{\theta}{2}}{2} + \frac{1 - \cos \frac{\theta}{2}}{2} (1 - 2 \hat{n}_{i, \uparrow}) (1 - 2 \hat{n}_{i+1, \uparrow})\; \\
    & - \sin \frac{\theta}{2} \Big[ \hat{c}_{i, \uparrow}  (1-2\hat{n}_{i, \downarrow}) \hat{c}^{\dagger}_{i+1, \uparrow} + \hat{c}^{\dagger}_{i, \uparrow} (1-2\hat{n}_{i, \downarrow}) \hat{c}_{i+1, \uparrow} \Big].
    \end{aligned}
\label{eq:D_theta_fermion}
\end{equation}
If $\ket{i, \uparrow}$ and $\ket{i + 1, \uparrow}$ are adjacent instead, the $Z_{i, \downarrow} \equiv 1 - 2 \hat{n}_{i, \downarrow}$ operators are absent from the spin-$\uparrow$ hopping terms, which gives rise to a phase shift of $-1$ between the original basis states with only singly-occupied sites and the newly added ones with empty and doubly-occupied sites. This minus sign inhibits any improvement of the overlap with the exact ground state.

Fig. \ref{fig3} shows the improvement of the fidelity relative to the Fermi-Hubbard model ground state of the fermionic version of the Heisenberg ground state --- prepared via the RVB-inspired ansatz with the minimum number of layers to achieve a fidelity of at least $0.99$ (see Fig. \ref{fig2}) --- upon applying this layer with a single parameter $\theta$. Remarkably, the convex $\theta \textrm{ vs. } \frac{U}{t}$ curves (shown in insets) obtained for chains and ladders of different sizes match almost perfectly (see Supp. Mat. \cite{SuppMatRef}), so this parameter does not have to be optimized through a hybrid scheme. In fact, the results presented in Fig. \ref{fig3} were obtained using the optimized $\theta$ from the simulations with the $10$-site lattices of the same geometry. Importantly, the improved fermionic Heisenberg state outperforms the noninteracting ground state for $\frac{U}{t} \gtrsim 4$ --- well within the most relevant regime of the Fermi-Hubbard model --- and even goes beyond the Gutzwiller wave function for a larger $\frac{U}{t}$.

In summary, we have developed a scheme to prepare an educated guess of the ground state of the Fermi-Hubbard model that may be adopted as the initial state on NISQ hardware. Further developments that we anticipate include simulations on non-bipartite lattices --- possibly modifying the reference state of the RVB-inspired ansatz to probe the right symmetry subspace (e.g., replacing a valence bond with a triplet to set $S_{\textrm{Total}} = 1$) ---, the consideration of the spin-$\frac{1}{2}$ model that includes $\mathcal{O}(t^4/U^3)$ ring-exchange terms in the VQE simulation, the exploration of analytical results to bypass the VQE simulation altogether --- e.g., Bethe ansatz states~\cite{Bethe31, VanDyke21, Sopena22} in 1D, Majumdar-Ghosh~\cite{MajumdarGhosh69a, MajumdarGhosh69b} and Shastry-Sutherland~\cite{ShastrySutherland81, Corboz13} states upon adding next-nearest-neighbor hopping terms in 1D and 2D, respectively --- and the extension of the scheme beyond half-filling --- e.g., replacing the Heisenberg model with the t-J model~\cite{Auerbach94} or changing the particle number through a chemical potential term. 

{\em Acknowledgments.} B.M. acknowledges financial support from Funda\c{c}\~{a}o para a Ci\^{e}ncia e a Tecnologia (FCT) --- Portugal through the Ph.D. scholarship No. SFRH/BD/08444/2020. 
J.F.R.  acknowledges financial support from
FCT (Grant No. PTDC/FIS-MAC/2045/2021),
Swiss National Science Foundation (SNSF) Sinergia (Grant Pimag),
Generalitat Valenciana funding Prometeo2021/017 and MFA/2022/045, 
and funding from MICIIN-Spain (Grant No. PID2019-109539GB-C41).

\newpage

\onecolumngrid


\begin{center}
  \textbf{\large{Supplemental Material for\\ ``From Heisenberg to Hubbard: An initial state for the \\ shallow quantum simulation of correlated electrons''}}\\[.2cm]
\end{center}

\setcounter{equation}{0}
\setcounter{figure}{0}
\setcounter{table}{0}
\renewcommand{\theequation}{S\arabic{equation}}
\renewcommand{\thefigure}{S\arabic{figure}}

\section{Preparing ground state of spin-$\frac{1}{2}$ Heisenberg model via VQE}

\subsection{The RVB-inspired ansatz}

The first part of the quantum scheme introduced in the main text involves the preparation of the ground state of the antiferromagnetic spin-$\frac{1}{2}$ Heisenberg model~\cite{Heisenberg28} via VQE~\cite{Peruzzo14} using the RVB-inspired ansatz~\cite{Seki20}. The basis gate decomposition of the building block of this ansatz, the so-called $\textrm{eSWAP}(\theta)$~\cite{Seki20}, is:
\begin{figure}[h]
\centering
\includegraphics[width=\linewidth]{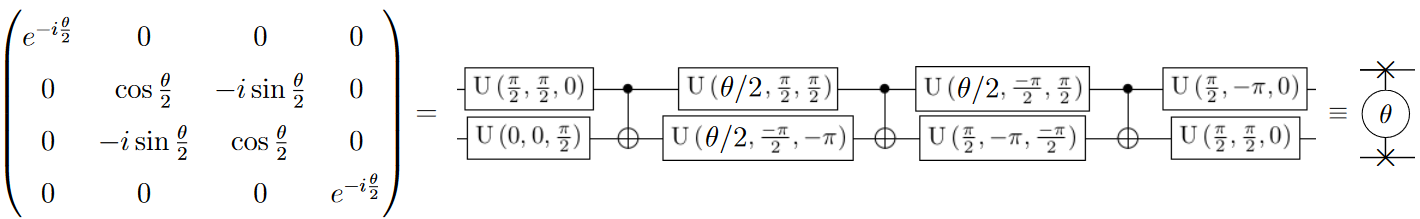}
\end{figure}

\noindent where $U(\theta, \phi, \lambda)$ is the general single-qubit operation~\cite{OpenQASM17}. Importantly, the eSWAP is SU(2)-invariant. An intuitive explanation for this property is the fact that angular momentum addition is commutative \cite{SakuraiNapolitano17}, so any function of the SWAP operation (including the $\textrm{eSWAP}(\theta) = e^{-i \frac{\theta}{2} \textrm{SWAP}}$) must commute with $\hat{S}^{i}_{\textrm{Total}}, i = x, y, z$. Hence, a circuit with only eSWAPs leaves the total spin $S_{\textrm{Total}}$ and its component along the quantization axis $S^{z}_{\textrm{Total}}$ unchanged.

\begin{figure}[t]
\centering
\includegraphics[width=\linewidth]{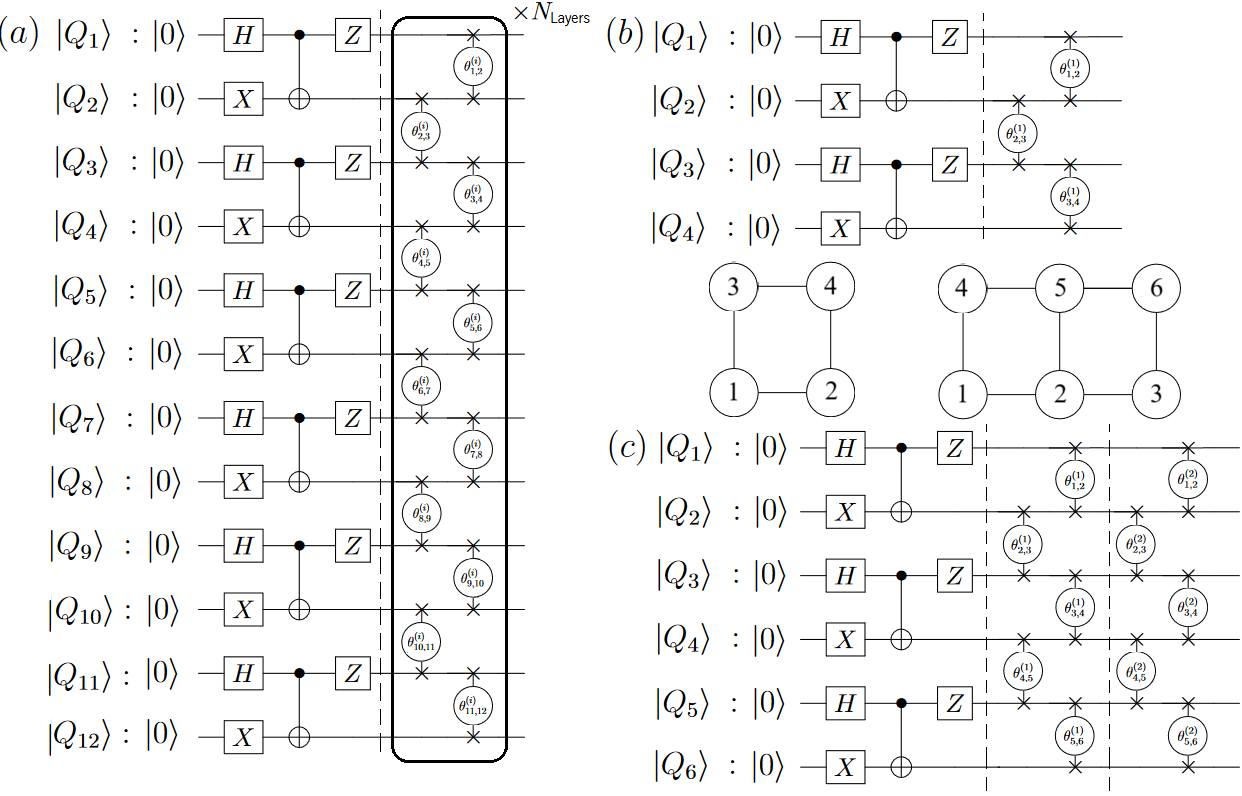}
\caption{(a) Scheme of RVB-inspired ansatz \cite{Seki20} to prepare ground state of antiferromagnetic spin-$\frac{1}{2}$ Heisenberg model \cite{Heisenberg28} via VQE \cite{Peruzzo14} for a lattice with $N = 12$ sites. The layer to the left of the barrier (dashed line) prepares the initial state corresponding to a product state of valence bonds $\frac{\ket{\uparrow \downarrow} - \ket{\downarrow \uparrow}}{\sqrt{2}}$ at pairs of adjacent qubits $(Q_{2j-1}, Q_{2j})$ with $j = 1, 2, ..., \frac{N}{2}$ for even $N$. On top of this initial state multiple parameterized layers are applied. A single layer (shown inside solid-line box) involves an eSWAP at every pair of adjacent qubits $(Q_i,Q_j)$ in a linearly connected architecture. Specifically, a layer of $\frac{N}{2}-1$ eSWAPs acting on the even-odd pairs $\{ (Q_{2j}, Q_{2j+1}) \}_{j = 1}^{\frac{N}{2}-1}$ is executed, followed by another layer of $\frac{N}{2}$ eSWAPs acting on the odd-even pairs $\{ (Q_{2j-1}, Q_{2j}) \}_{j = 1}^{\frac{N}{2}}$. Hence, the number of free parameters per layer is $N-1$, resulting in a total of $N_{\textrm{Layers}}(N-1)$ parameters. Given the decomposition of the eSWAP shown in the previous page, the circuit depth is of $6 N_{\textrm{Layers}} + 1$ CNOTs (including the preparation of the product state of valence bonds) for any scenario of qubit connectivity. (b,c) Examples of application of RVB-inspired ansatz to encode exact ground state of antiferromagnetic spin-$\frac{1}{2}$ Heisenberg model on $2 \times 2$ (b) and $3 \times 2$ (c) lattices. Sites are ordered as shown in the diagrams, with qubit $Q_i$ encoding the spin-$\frac{1}{2}$ at site $i$. For the $2 \times 2$ lattice (b), a single layer allows to prepare the exact ground state, with parameter values $(\theta^{(1)}_{1,2}, \theta^{(1)}_{2,3}, \theta^{(1)}_{3,4}) = (0, -\arccos(1/3), \arccos(1/3))$. As for the $3 \times 2$ lattice (c), two layers of the RVB-inspired ansatz are needed to initialize the true ground state, with parameters (in radians) $(\theta^{(1)}_{1,2}, \theta^{(1)}_{2,3}, \theta^{(1)}_{3,4}, \theta^{(1)}_{4,5}, \theta^{(1)}_{5,6}, \theta^{(2)}_{1,2}, \theta^{(2)}_{2,3}, \theta^{(2)}_{3,4}, \theta^{(2)}_{4,5}, \theta^{(2)}_{5,6}) = (4.76343956, 0.94395129, 4.42318361, 1.24182304, 6.90086599, 4.40573414, 1.85758957,$  $3.86482876, 2.64778842, 5.85341176)$.}
\label{fig:RVB_inspired_ansatz}
\end{figure}

For the sake of clarity, Fig. \ref{fig:RVB_inspired_ansatz} shows a scheme of the quantum circuit that realizes the RVB-inspired ansatz for a $12$-site lattice (a), as well as two examples of the application of this ansatz to prepare the exact ground state of the antiferromagnetic spin-$\frac{1}{2}$ Heisenberg model on $2 \times 2$ (b) and $3 \times 2$ (c) lattices. Importantly, the ordering of the sites shown in the diagrams of the lattices --- i.e., sites are numbered along the rows --- is followed for all $L \times 2$ ladder geometries considered in this work. Given the basis gate decomposition of the eSWAP shown above, the depth of the circuit that prepares the RVB-inspired ansatz that explores the $S_{\textrm{Total}} = 0$ subspace is $6 N_{\textrm{Layers}} + 1$, including the $1$-CNOT-depth layer associated with the initialization of the product state of valence bonds. This circuit depth is valid for any scenario of qubit connectivity, as no rerouting of qubits is required even in the most restrictive case of linear qubit connectivity.

The simulations performed in this work involved balanced bipartite lattices with an even number of sites, in which case the Lieb-Mattis theorem~\cite{LiebMattis62} determines that the exact ground state of the antiferromagnetic spin-$\frac{1}{2}$ Heisenberg model has $S_{\textrm{Total}} = 0$. For imbalanced bipartite lattices or non-bipartite lattices, the good quantum numbers $(S_{\textrm{Total}}, S^{z}_{\textrm{Total}})$ may not be $(0,0)$. To explore the right symmetry subspace, the reference state of the RVB-inspired ansatz must be changed accordingly. For example, to consider the $(1,0)$, $(1,1)$ or $(1,-1)$ sectors, it suffices to replace one of the valence bonds $\frac{\ket{\uparrow \downarrow} - \ket{\downarrow \uparrow} }{\sqrt{2}}$ with the appropriate triplet state $\frac{\ket{\uparrow \downarrow} + \ket{\downarrow \uparrow} }{\sqrt{2}}$, $\ket{\uparrow \uparrow}$ or $\ket{\downarrow \downarrow}$. Likewise, for an odd number of lattice sites, we may have an unpaired spin-$\frac{1}{2}$ in the reference state.

\subsection{Optimization of parameters}

The ground state of the antiferromagnetic spin-$\frac{1}{2}$ Heisenberg model was determined via VQE \cite{Peruzzo14} using the RVB-inspired ansatz defined in the previous section. The parameters were optimized by minimizing the energy using the \texttt{scipy.optimize.minimize} function \cite{scipyminimize}. Since all parameters were bounded to the $[0, 2\pi]$ interval, this function implemented the Sequential Least Squares Programming (SLSQP) method. For a sufficiently large number of sites $N$, the optimization landscape became complex enough for the optimizer to become trapped in local minima for multiple sets of initial conditions. A two-fold strategy was adopted to overcome this issue. First, after having optimized the $(N_{\textrm{Layers}}-1)$-layer ansatz, which resulted in a set of $(N_{\textrm{Layers}}-1)(N-1)$ parameter values $\vec{\theta}_{N_{\textrm{Layers}}-1}$, the initial conditions for the next step corresponding to the optimization of the $N_{\textrm{Layers}}$-layer ansatz were set as $\vec{\theta}^{\textrm{init}}_{N_{\textrm{Layers}}} = (\vec{\theta}_{N_{\textrm{Layers}}-1}, \vec{0})$, where $\vec{0}$ is a $(N-1)$-dimensional vector with all entries set to zero. The key point is that a layer of the RVB-inspired ansatz with all parameters set to zero amounts to the identity, so this corresponds to a layer-by-layer strategy where the outcome of one iteration is used as the starting point of the next. Second, to avoid being trapped in this initial state (which is often associated with a local minimum of the optimization landscape), a random value in the range $[-2\eta \pi, 2\eta \pi]$ --- with $\eta \in [0, 1/2]$ a simulation parameter --- was added to every entry of $\vec{\theta}^{\textrm{init}}_{N_{\textrm{Layers}}}$. In practice, five different values of the noise parameter, $\eta \in \{ 0.1, 0.2, 0.3, 0.4, 0.5 \}$, were considered, with $N_{\textrm{reps}} = 10$ repetitions performed for each. Hence, for each lattice geometry and every $N_{\textrm{Layers}}$, a total of $50$ trials of the optimization process were carried out, each involving the optimization of all $N_{\textrm{Layers}}(N-1)$ parameters.

\section{Converting multi-spin-$\frac{1}{2}$ wave function into fermionic version}

\begin{figure}[b]
\centering
\includegraphics[width=\linewidth]{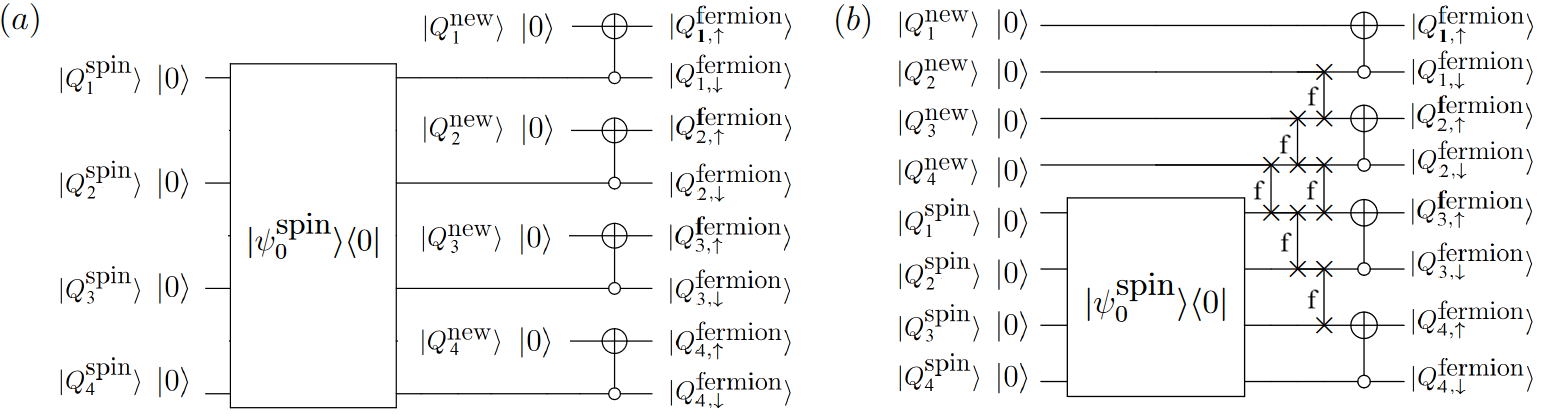}
\caption{Quantum circuit to transform $N$-spin-$\frac{1}{2}$ wave function $\ket{\psi^{\textrm{spin}}_{0}}$ into spin-$\frac{1}{2}$ fermionic wave function on lattice with $N = 4$ sites at half-filling with qubits ordered by site for all-to-all qubit connectivity (a) and linear qubit connectivity (b). In the former case, the conversion layer has just $1$ CNOT of depth. In the latter case, the qubits must be ordered by site instead of spin through a network of fermionic SWAPs \cite{Verstraete09}.}
\label{fig:spin_to_fermion_circuit_site_order}
\end{figure}

The conversion of an arbitrary multi-spin-$\frac{1}{2}$ wave function into its fermionic version at half-filling was explained in the main text. The purpose of this section of the Supplemental Material is to clarify this process when qubit connectivity constraints are present and, if the qubits are ordered by spin, when the given multi-spin-$\frac{1}{2}$ state does not have a well-defined $S^{z}_{\textrm{Total}}$.

First, let us assume the qubit are ordered by site, as in $\ket{1\uparrow} \ket{1\downarrow} \ket{2\uparrow} \ket{2\downarrow} ... \ket{N\uparrow} \ket{N\downarrow}$. In such case, the amplitudes of $\ket{\psi^{\textrm{fermion}}}$ are exactly equal to those of $\ket{\psi^{\textrm{spin}}}$. To understand why this is so, it is instructive to consider how the Hilbert space for multiple spins-$\frac{1}{2}$ is constructed out of the local Hilbert space of a single one. For just one spin-$\frac{1}{2}$, the Hilbert space is $\textrm{span} \{ \ket{\uparrow} \leftrightarrow \hat{c}^{\dagger}_{\uparrow} \ket{\Omega}, \ket{\downarrow} \leftrightarrow \hat{c}^{\dagger}_{\downarrow} \ket{\Omega} \}$, where the corresponding fermionic representation for a single site at half-filling is also displayed, with $\ket{\Omega}$ the vacuum state. Once multiple spins-$\frac{1}{2}$ are considered, the full Hilbert space is generated through the tensor product of the individual two-dimensional spaces at each site. Hence, for example, a basis state of the form $\ket{\uparrow \downarrow \downarrow \uparrow \downarrow \uparrow}$ corresponds explicitly to $\ket{\uparrow}_{1} \otimes \ket{\downarrow}_{2} \otimes \ket{\downarrow}_{3} \otimes \ket{\uparrow}_{4} \otimes \ket{\downarrow}_{5} \otimes \ket{\uparrow}_{6}$, the fermionic version of which is naturally ordered by site as well: $\hat{c}^{\dagger}_{1, \uparrow} \hat{c}^{\dagger}_{2, \downarrow} \hat{c}^{\dagger}_{3, \downarrow} \hat{c}^{\dagger}_{4, \uparrow} \hat{c}^{\dagger}_{5, \downarrow} \hat{c}^{\dagger}_{6, \uparrow} \ket{\Omega}$. As a result, ordering the $2N$ qubits by site, the conversion of $\ket{\psi^{\textrm{spin}}}$ into $\ket{\psi^{\textrm{fermion}}}$ merely amounts to repeating the application of the 1-CNOT two-qubit circuit shown in Eq. (4) in the main text at every pair of qubits representing a lattice site. Fig. \ref{fig:spin_to_fermion_circuit_site_order} shows the resulting quantum circuits for all-to-all (a) and linear (b) qubit connectivity. In the latter case, it is convenient to have the $N$ qubits on which $\ket{\psi^{\textrm{spin}}}$ is prepared next to one another initially; these qubits are then rerouted through a network of fermionic SWAPs~\cite{Verstraete09}. Note that, although at first glance one might think that conventional SWAPs, rather than fermionic ones, would have to be employed in this rerouting of the qubits since at this point the qubits do not actually represent the occupations of fermionic orbitals, in practice both conventional and fermionic SWAPs produce exactly the same effect because all ancillas are in state $\ket{0}$, so the latter are preferred as they take only $2$ CNOTs~\cite{CruzMurta23} instead of $3$~\cite{Nielsen00}.

\begin{figure}[t]
\centering
\includegraphics[width=\linewidth]{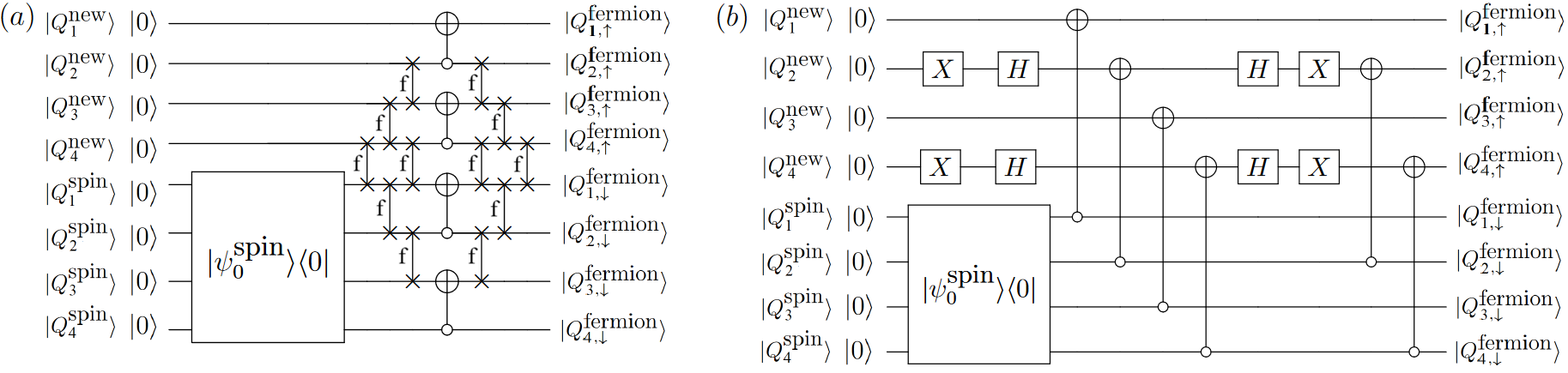}
\caption{Quantum circuit to transform $N$-spin-$\frac{1}{2}$ wave function $\ket{\psi^{\textrm{spin}}_{0}}$ into spin-$\frac{1}{2}$ fermionic wave function on lattice of $N = 4$ sites at half-filling with qubits ordered by spin for linear qubit connectivity (a) and all-to-all qubit connectivity (b). The latter case results in a constant circuit depth overhead of just $2$ CNOTs but only works if $S^{z}_{\textrm{Total}}$ is a good quantum number of $\ket{\psi_{0}^{\textrm{spin}}}$. If this is not the case, then the $\mathcal{O}(N)$-depth circuit in (a) must be considered even if there are no connectivity constraints.}
\label{fig:spin_to_fermion_circuit_spin_order}
\end{figure}

If we wish to have the $2N$ qubits ordered by spin instead of site, as in $\ket{1\uparrow} \ket{2\uparrow} ... \ket{N\uparrow} \ket{1\downarrow} \ket{2\downarrow} ... \ket{N\downarrow}$, although the absolute values of all amplitudes of $\ket{\psi^{\textrm{spin}}}$ remain unchanged in $\ket{\psi^{\textrm{fermion}}}$, some may have their signs changed. Returning to the example from the previous paragraph, upon commuting the creation operators in the basis state $\ket{\uparrow \downarrow \downarrow \uparrow \downarrow \uparrow} \equiv \hat{c}^{\dagger}_{1, \uparrow} \hat{c}^{\dagger}_{2, \downarrow} \hat{c}^{\dagger}_{3, \downarrow} \hat{c}^{\dagger}_{4, \uparrow} \hat{c}^{\dagger}_{5, \downarrow} \hat{c}^{\dagger}_{6, \uparrow} \ket{\Omega}$, we obtain $-\hat{c}^{\dagger}_{1, \uparrow} \hat{c}^{\dagger}_{4, \uparrow} \hat{c}^{\dagger}_{6, \uparrow} \hat{c}^{\dagger}_{2, \downarrow} \hat{c}^{\dagger}_{3, \downarrow} \hat{c}^{\dagger}_{5, \downarrow} \ket{\Omega}$, so an extra minus sign must be applied to its amplitude. The conceptually simplest and most general way of keeping track of these fermionic signs is to apply the inverse of the network of fermionic SWAPs shown in Fig. \ref{fig:spin_to_fermion_circuit_site_order}(b) to transform the site-ordered fermionic wave function back into its spin-ordered version, as shown in Fig. \ref{fig:spin_to_fermion_circuit_spin_order}(a). This is, in fact, the shallowest solution when one is restricted to linear qubit connectivity. However, in the absence of connectivity constraints, it is inconvenient to have to turn from spin order to site order and then back to spin order when we already have the qubits ordered by spin in the first place. It is, in fact, possible to avoid this when $S^{z}_{\textrm{Total}} = \frac{N^{\uparrow} - N^{\downarrow}}{2}$ is a good quantum number --- i.e., all basis states with nonzero amplitudes have the same numbers of spins-$\uparrow$, $N^{\uparrow}$, and spins-$\downarrow$, $N^{\downarrow}$, with $N^{\uparrow} + N^{\downarrow} = N$ in order to be at half-filling (which is required to have a well-defined spin degree of freedom at each site). In such case, at every other site the two-qubit operation stated in Eq. (4) in the main text is simply replaced by Eq. (5), as illustrated in Fig. \ref{fig:spin_to_fermion_circuit_spin_order}(b) for $N = 4$ sites. The depth overhead of this layer that converts $\ket{\psi^{\textrm{spin}}}$ into $\ket{\psi^{\textrm{fermion}}}$ is of just $2$ CNOTs, which is independent of the lattice size, as for the site-ordered case. This approach is valid for an arbitrary state with any value of $S^{z}_{\textrm{Total}}$, as confirmed numerically for the ground states of the Heisenberg model on the chain and ladder geometries considered in this work and for randomly generated states with an even number of lattice sites up to $N = 8$. If $\ket{\psi^{\textrm{spin}}}$ includes terms with different values of the magnetization along the quantization axis, then the default method shown in Fig. \ref{fig:spin_to_fermion_circuit_spin_order}(a) must be adopted even with unrestricted qubit connectivity. 

In the end, given a $N$-qubit subcircuit (such as the RVB-inspired ansatz circuit with optimized parameters) to initialize an approximation to the exact ground state of the antiferromagnetic spin-$\frac{1}{2}$ Heisenberg model $\ket{\psi^{\textrm{spin}}_{0}}$ with infidelity $\epsilon$, an approximation $\ket{\psi^{\textrm{fermion}}_{0}}$ of the exact ground state of the Fermi-Hubbard model in the strongly-interacting limit $\frac{U}{t} \to \infty$ is prepared on a $2N$-qubit register with the same infidelity $\epsilon$, ignoring $\mathcal{O}(t/U)$ corrections. Note that this state has the expected good quantum numbers, namely the correct total particle number $N_{\textrm{particles}} = N$, total spin $S_{\textrm{Total}} = 0$ and its component along the quantization axis $S^{z}_{\textrm{Total}} = 0$.

\section{Improving overlap with Fermi-Hubbard ground state at finite $\frac{U}{t}$}

\subsection{Structure of the ansatz that adds doublon-holon pairs}

Having prepared the fermionic version of the ground state of the spin-$\frac{1}{2}$ Heisenberg model $\ket{\psi^{(U/t \to \infty)}_{0}}$, which corresponds to the exact ground state of the Fermi-Hubbard model at half-filling in the $\frac{U}{t} \to \infty$ limit, we now wish to improve its overlap with the exact ground state at a finite $\frac{U}{t}$. To accomplish this, a layer of two-qubit operations
\begin{equation}
    \begin{pmatrix}
    1 & 0 & 0 & 0 \\
    0 & \cos \frac{\theta}{2} & - \sin \frac{\theta}{2} & 0 \\
    0 & \sin \frac{\theta}{2} & \cos \frac{\theta}{2} & 0 \\
    0 & 0 & 0 & 1
    \end{pmatrix} = 
    \raisebox{-0.42\totalheight}{
    \begin{tikzpicture}
    \node[scale=0.8]{
    \begin{tikzcd}[row sep={1.0cm,between origins}, column sep=0.2cm]
    & \gate{U(\frac{\pi}{2},-\pi, \frac{5 \pi}{8})} & \ctrl{1} & \gate{U(\frac{\theta}{2}, 0, 0)} & \ctrl{1} & \gate{U(\frac{\pi}{2}, -\frac{5 \pi}{8}, -\pi)} & \qw \\
    & \gate{U(0,0,\frac{5\pi}{8})} & \targ{} & \gate{U(\pi - \frac{\theta}{2}, 0, 0)} & \targ{} & \gate{\pi, 0, -\frac{3 \pi}{8}} & \qw 
    \end{tikzcd}
    };
    \end{tikzpicture}} \equiv
    \raisebox{-0.42\totalheight}{
    \begin{tikzpicture}
    \node[scale=0.8]{
    \begin{tikzcd}[row sep={1.0cm,between origins}, column sep=0.3cm]
    & \gate[2]{D(\theta)} & \qw \\
    &  & \qw 
    \end{tikzcd}
    };
    \end{tikzpicture}}
\label{eq:doublon_holon_element_matrix}
\end{equation}
is applied to $\ket{\psi^{(U/t \to \infty)}_{0}}$ at the qubits encoding the spin-$\uparrow$ orbitals of adjacent odd-even pairs of sites $(1,2), (3,4), ..., (N-1, N)$. All $\frac{N}{2}$ two-qubit operations have the same parameter $\theta$. In line with Eq. (6) in the main text, a nonzero value of $\theta$ causes $D(\theta)$ to introduce pairs of empty and doubly-occupied sites by promoting the hopping of spin-$\uparrow$ electrons between the two sites in question. Hence, we expect that, as $\frac{U}{t}$ decreases, the doublon-holon pairs become less energetically costly, so the weight of basis states carrying empty and doubly-occupied sites in the exact ground state of the Fermi-Hubbard model should increase, which is reflected by an increase of the optimal value of the free parameter $\theta$ found by minimizing the energy of this ansatz.

\begin{figure}[t]
\centering
\includegraphics[width=\linewidth]{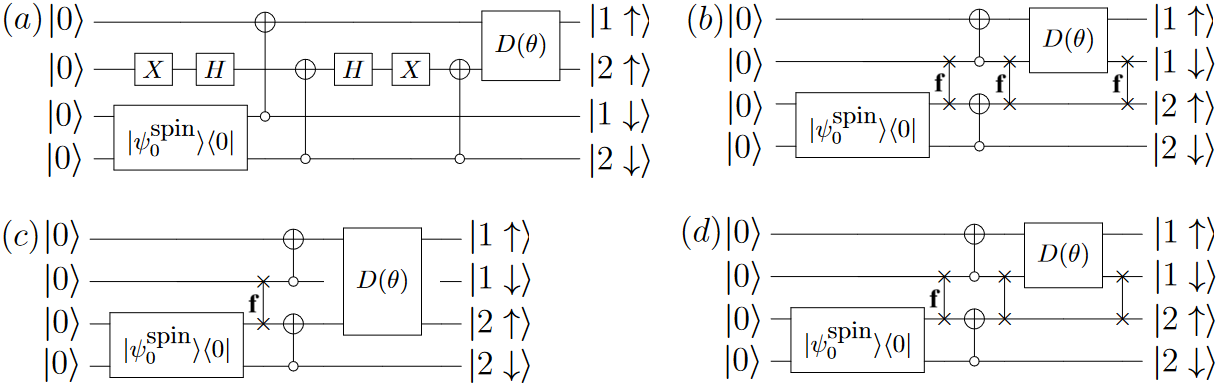}
\caption{Unsuccessful (a-b) and successful (c-d) implementations of heuristic parameterized layer to improve overlap of fermionic version of Heisenberg ground state with exact ground state of Fermi-Hubbard dimer at finite $\frac{U}{t}$. $D(\theta)$ is defined in Eq. (\ref{eq:doublon_holon_element_matrix}). Circuits in (a) and (c) assume all-to-all connectivity, while those in (b) and (d) are restricted to linear connectivity. The improvement of the overlap with the exact ground state at finite $\frac{U}{t}$ for an optimized parameter $\theta$ only occurs when the qubit that encodes the occupation of the spin-$\downarrow$ orbital at site $1$ is between the two qubits encoding the spin-$\uparrow$ orbitals on which $D(\theta)$ acts nontrivially. Notice the subtle difference between (b) and (d): in the former (unsuccessful) case, the moving qubit is rerouted with fermionic SWAPs, while in the later (successful) case the qubit rerouting is performed with conventional SWAPs.}
\label{fig:D_theta_wrong_and_right_implementations_dimer}
\end{figure}

Upon applying this ansatz, it turns out that having the $\ket{i, \downarrow}$ qubit between the $(\ket{i, \uparrow}, \ket{i+1, \uparrow})$ qubits on which $D(\theta)$ acts nontrivially is decisive to obtain the expected improvement of the overlap with the exact ground state at finite $\frac{U}{t}$ with respect to $\ket{\psi^{(U/t \to \infty)}_0}$. This is made clear by considering the Fermi-Hubbard dimer (i.e., a $2 \times 1$ lattice). In this case, $\ket{\psi^{(U/t \to \infty)}_0} = \frac{1}{\sqrt{2}} (\ket{\uparrow, \downarrow} - \ket{\downarrow, \uparrow})$ (i.e., $N_{\textrm{Layers}} = 0$ in the RVB-inspired ansatz). The exact ground state at an arbitrary interaction strength $\frac{U}{t}$ is given by 
\begin{equation}
    \ket{\psi_{\textrm{exact}}} = \mathcal{N} ( \ket{\uparrow \downarrow, 0} + \alpha \ket{\uparrow, \downarrow} - \alpha \ket{\downarrow, \uparrow} + \ket{0, \uparrow \downarrow} ),
\label{eq:FHM_dimer_gs}
\end{equation}
where $\alpha = \frac{U + \sqrt{U^2 + 16 t^2}}{4t}$, $\mathcal{N} = \frac{1}{\sqrt{2}\sqrt{1 + \alpha^2}}$ normalizes the wave function, and the basis states are ordered by site instead of spin. However, upon applying $D(\theta)$ at the pair of \textit{adjacent} qubits $(\ket{1\uparrow}, \ket{2\uparrow})$ of $\ket{\psi^{(U/t \to \infty)}_0}$ (see Fig. \ref{fig:D_theta_wrong_and_right_implementations_dimer}(a)-(b)), we obtain the following state
\begin{equation}
    D(\theta) \ket{\psi^{(U/t \to \infty)}_0} = \cos \frac{\theta}{2} \left( \frac{\ket{\uparrow, \downarrow} - \ket{\downarrow, \uparrow}}{\sqrt{2}} \right) - \sin \frac{\theta}{2} \left( \frac{\ket{\uparrow \downarrow, 0} + \ket{0, \uparrow \downarrow}}{\sqrt{2}} \right). 
\end{equation}
Were it not for the relative phase factor of $-1 = e^{i \pi}$ between the left-hand-side part involving only half-filled sites and the right-hand-side part with doublon-holon pairs, this ansatz would give rise to an exact representation of the true ground state of the Fermi-Hubbard dimer at any $\frac{U}{t}$ for a suitable choice of $\theta$ (cf. Eq. (\ref{eq:FHM_dimer_gs})). To remove this minus sign, we can instead apply the same two-qubit operation $D(\theta)$ but with the qubit encoding the spin-$\downarrow$ orbital at site $1$ between the two qubits on which $D(\theta)$ acts nontrivially, as shown in Fig. \ref{fig:D_theta_wrong_and_right_implementations_dimer}(c)-(d). The only change this introduces is an extra $Z_{i, \downarrow} = 1 - 2 \hat{n}_{i, \downarrow}$ operator in the hopping terms of $D(\theta)$, as shown in Eq. (7) in the main text. 

The same behavior is observed for larger lattices: applying $D(\theta)$ to adjacent qubits encoding spin-$\uparrow$ orbitals leads to no improvement of the fidelity relative to the exact ground state, but introducing the qubit storing the occupation of the spin-$\downarrow$ orbital between them does allow for an enhancement of the overlap. Interestingly, while the Z strings arising from the anti-commutatitivity of fermionic operators are often regarded as inconvenient because they convert local fermionic operators into non-local qubit operations, here the $Z_{i \downarrow} = 1 - 2 \hat{n}_{i \downarrow}$ introduced in the spin-$\uparrow$ hopping terms (see Eq. (7) in the main text) turned out to be useful. 

\subsection{Additional numerical results for chains and ladders of multiple sizes}

\begin{figure}[t]
\centering
\includegraphics[width=\linewidth]{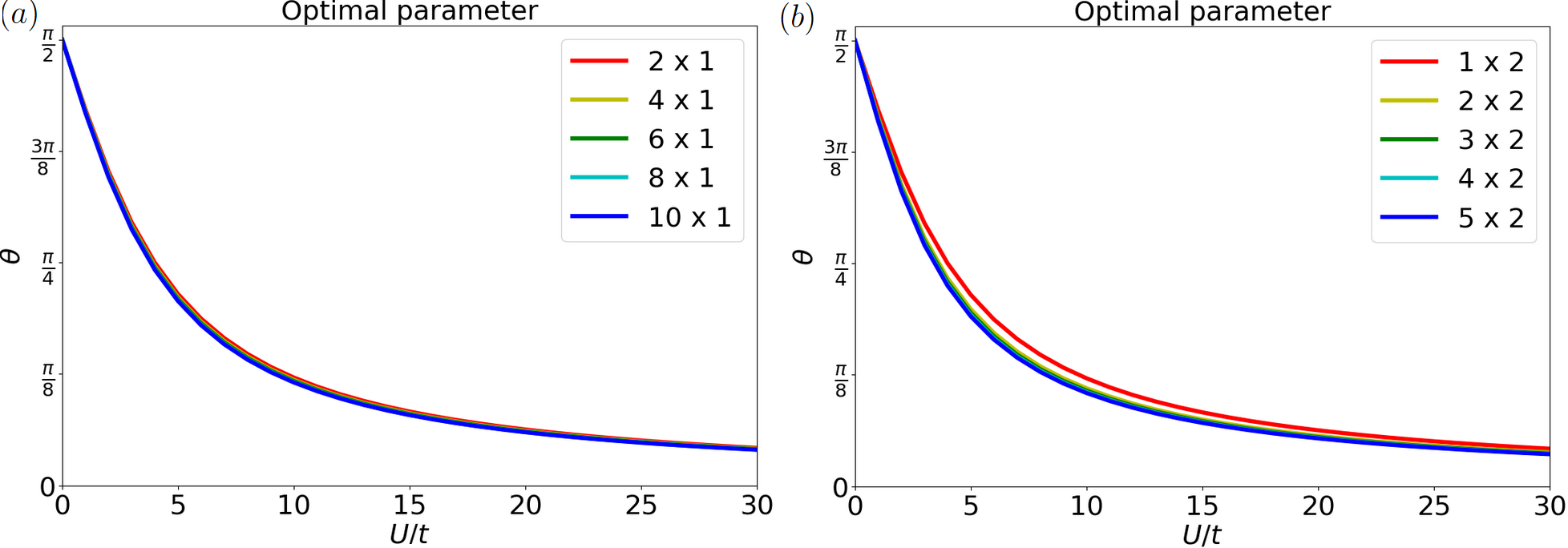}
\caption{Optimal value of free parameter $\theta$ of constant-depth layer that adds doublon-holon pairs to fermionic version of Heisenberg ground state for $L \times 1$ (a) and $L \times 2$ (b) lattices. $\theta$ decreases monotonically with $\frac{U}{t}$, resulting in a simple convex landscape. The $\theta \textrm{ vs. } \frac{U}{t}$ profiles for lattices with the same geometry but different sizes are nearly coincident.}
\label{figS5}
\end{figure}

Fig. \ref{figS5} shows the optimal value of the free parameter $\theta$ against the interaction strength for the chains and ladders of different sizes considered in this work. Not only is the $\theta \textrm{ vs. } \frac{U}{t}$ profile a simple convex curve for all geometries, but there is a nearly perfect matching of the quantitative values for lattices with the same geometry but different sizes. Hence, this parameter does not even require explicit optimization through a hybrid (quantum-classical) scheme: the optimal value for larger systems that cannot be simulated on conventional hardware can be simply obtained from the \textit{in silico} simulations of smaller systems.

\begin{figure}[h!]
\centering
\includegraphics[width=\linewidth]{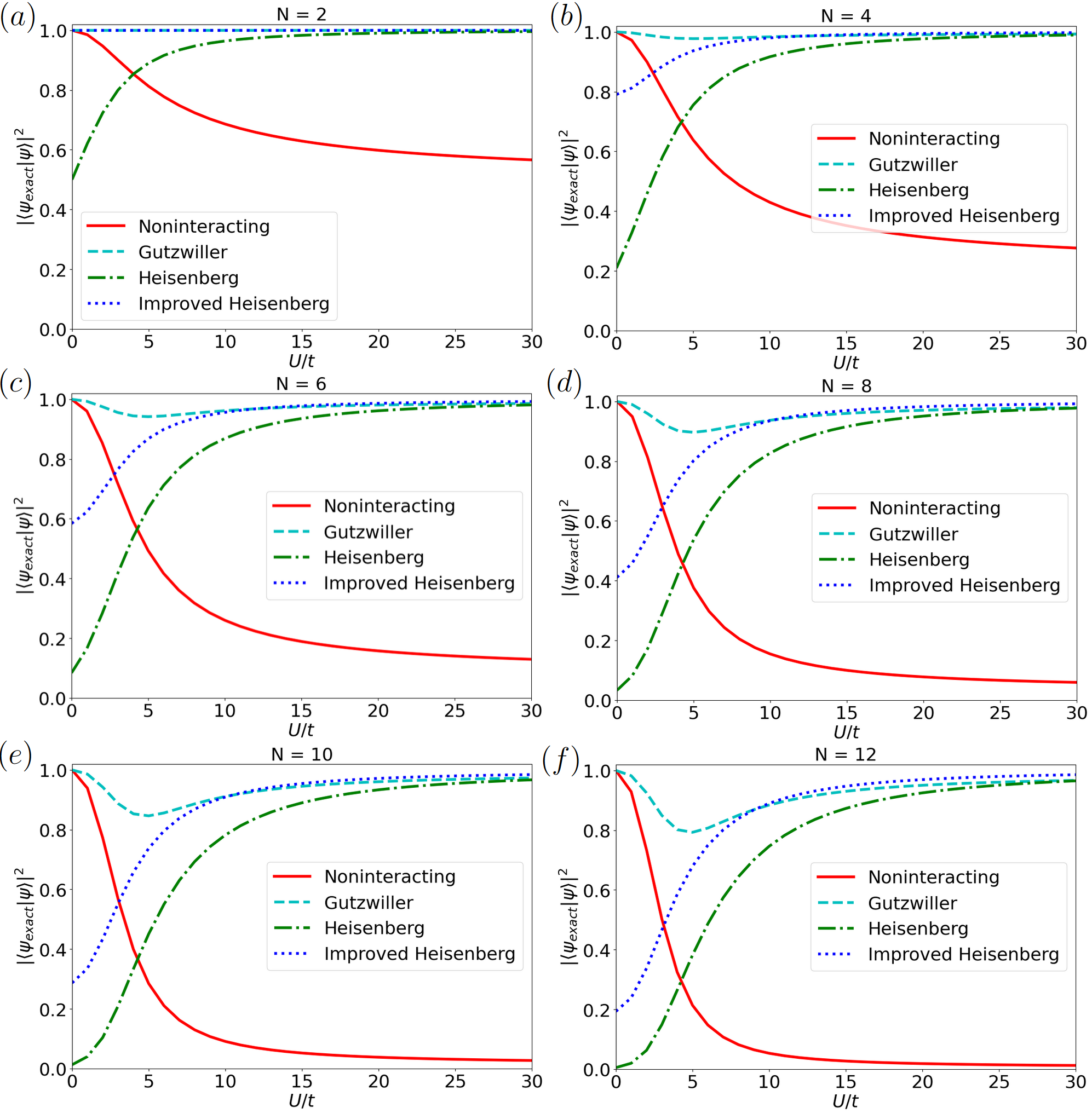}
\caption{Fidelity relative to exact ground state of noninteracting ground state (red solid line), Gutzwiller wave function (cyan dashed line), fermionic version of Heisenberg ground state (green dashed-dotted line) and its improved version (blue dotted line) for Fermi-Hubbard chains with $N = 2,4,6,8,10,12$ sites and open boundary conditions. The RVB-inspired ansatz with the minimum number of layers to achieve a fidelity of at least $0.99$ was used to initialize the Heisenberg ground state (see Fig. 2 in main text).}
\label{figS6}
\end{figure}

\begin{figure}
\centering
\includegraphics[width=\linewidth]{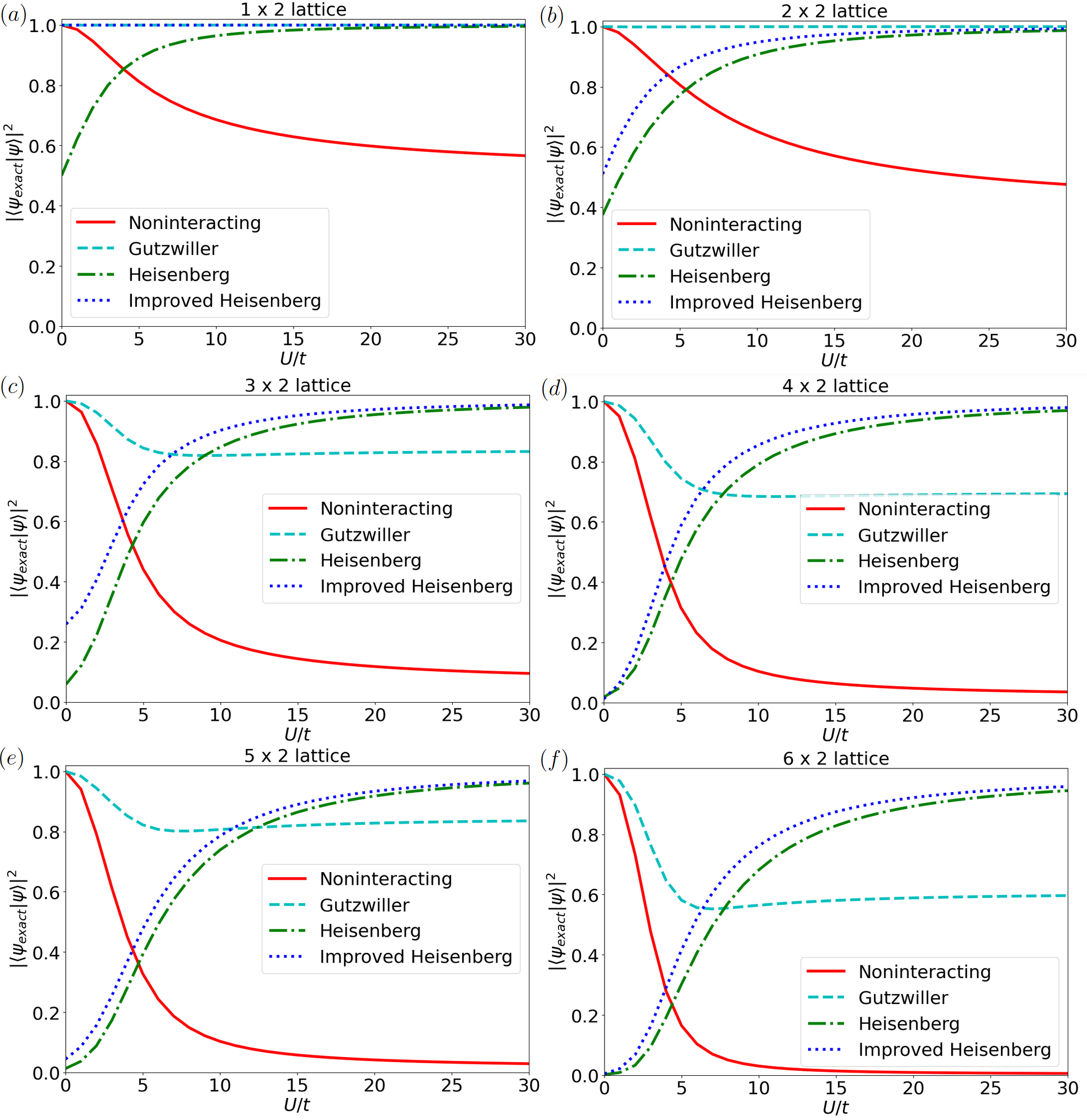}
\caption{Fidelity relative to exact ground state of noninteracting ground state (red solid line), Gutzwiller wave function (cyan dashed line), fermionic version of Heisenberg ground state (green dashed-dotted line) and its improved version (blue dotted line) for $L_x \times 2$ Fermi-Hubbard ladders, with $L_x = 1,2,3,4,5,6$ and open boundary conditions. The RVB-inspired ansatz with the minimum number of layers to achieve a fidelity of at least $0.99$ was used to initialize the Heisenberg ground state (see Fig. 2 in main text).}
\label{figS7}
\end{figure}

\begin{figure}
\centering
\includegraphics[width=\linewidth]{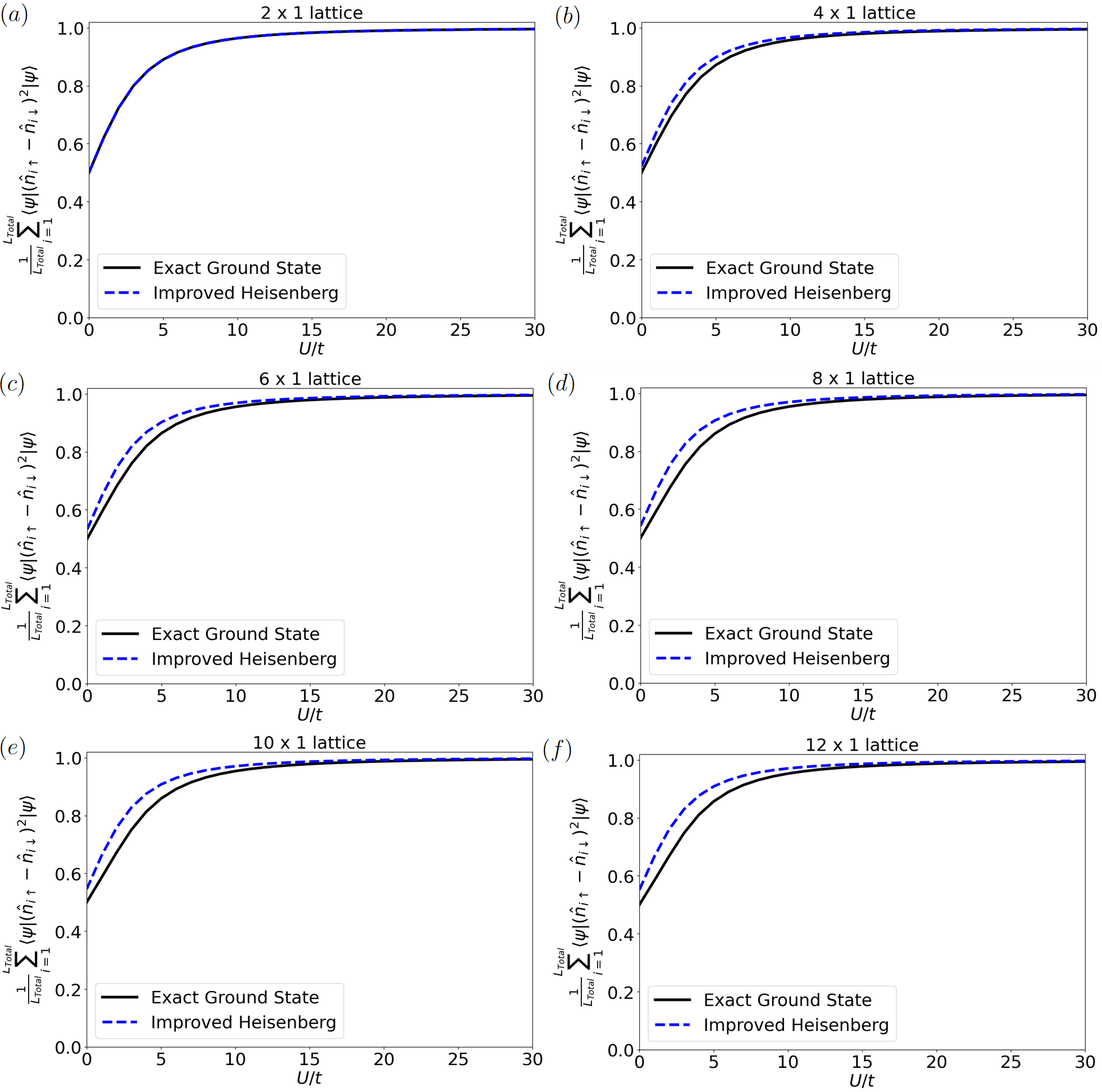}
\caption{Average across all lattice sites of expectation value of local moment $(\hat{n}_{i \uparrow} - \hat{n}_{i \downarrow})^2$, an indicator of double occupancy, for exact ground state of Fermi-Hubbard model and improved fermionic version of Heisenberg ground state for chains with $2, 4, 6, 8, 10, 12$ sites.}
\label{figS8}
\end{figure}

\begin{figure}
\centering
\includegraphics[width=\linewidth]{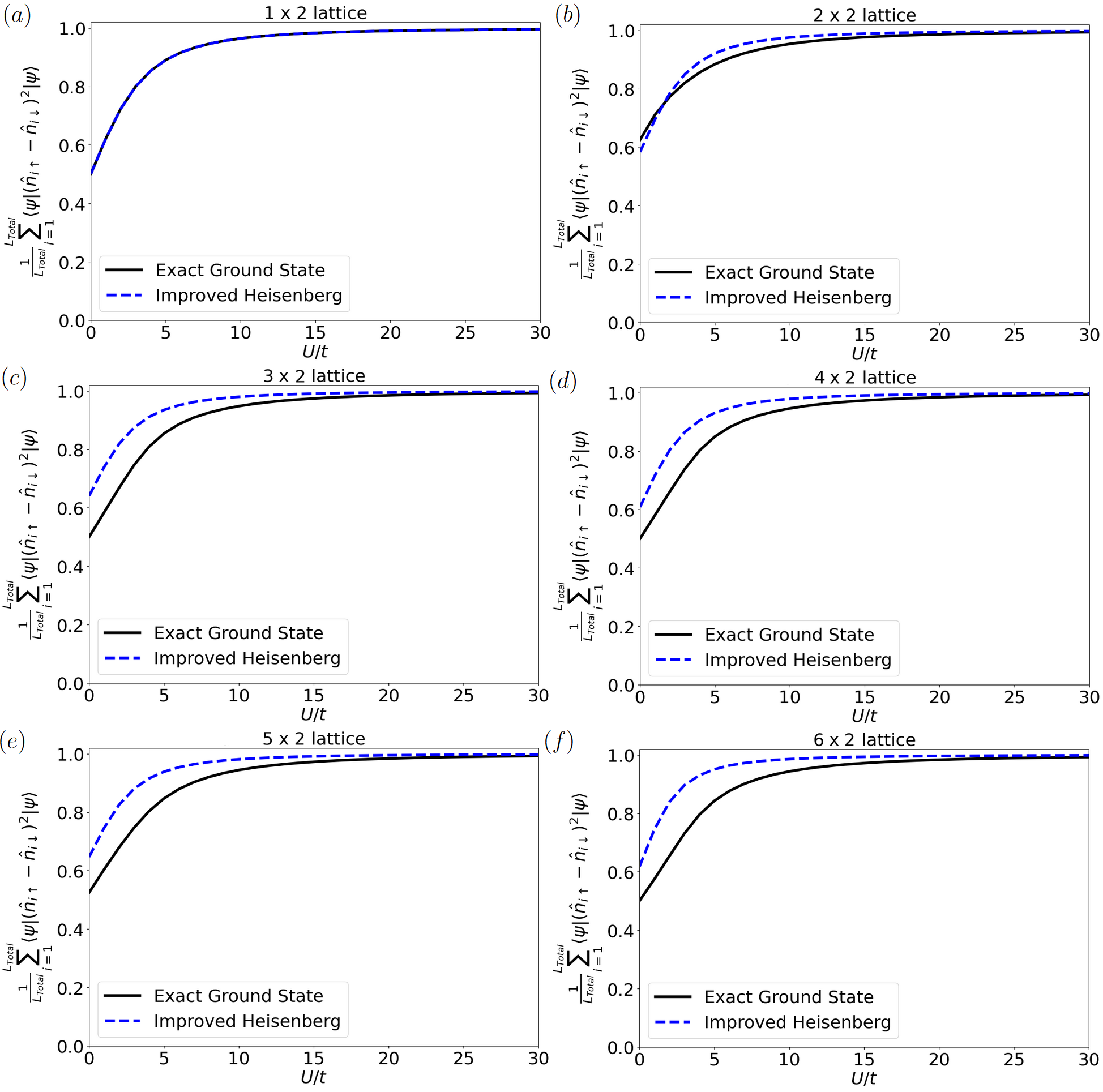}
\caption{Average across all lattice sites of expectation value of local moment $(\hat{n}_{i \uparrow} - \hat{n}_{i \downarrow})^2$, an indicator of double occupancy, for exact ground state of Fermi-Hubbard model and improved fermionic version of Heisenberg ground state for $L_x \times 2$ ladders with $L_x = 1, 2, 3, 4, 5, 6$.}
\label{figS9}
\end{figure}

Figs. \ref{figS6} and \ref{figS7} present numerical results analogous to those shown in Fig. 3 in the main text but for all sizes of chains and ladders considered in our work. Figs. \ref{figS8} and \ref{figS9} show the average over all lattice sites of the expectation value of the local moment, $(\hat{n}_{i \uparrow} - \hat{n}_{i \downarrow})^2 = \hat{n}_{i \uparrow} + \hat{n}_{i \downarrow} - 2 \hat{n}_{i \uparrow} \hat{n}_{i \downarrow}$, which is a measure of the number of doubly-occupied sites in the lattice. As expected, the rise in the optimal free parameter $\theta$ value as $\frac{U}{t}$ decreases is reflected in a greater weight of configurations with empty and doubly-occupied sites of the ansatz, which follows the same behavior observed in the exact ground state.

\bibliographystyle{apsrev4-2}
\bibliography{main}{}

\begin{thebibliography}{48}%
\makeatletter
\providecommand \@ifxundefined [1]{%
 \@ifx{#1\undefined}
}%
\providecommand \@ifnum [1]{%
 \ifnum #1\expandafter \@firstoftwo
 \else \expandafter \@secondoftwo
 \fi
}%
\providecommand \@ifx [1]{%
 \ifx #1\expandafter \@firstoftwo
 \else \expandafter \@secondoftwo
 \fi
}%
\providecommand \natexlab [1]{#1}%
\providecommand \enquote  [1]{``#1''}%
\providecommand \bibnamefont  [1]{#1}%
\providecommand \bibfnamefont [1]{#1}%
\providecommand \citenamefont [1]{#1}%
\providecommand \href@noop [0]{\@secondoftwo}%
\providecommand \href [0]{\begingroup \@sanitize@url \@href}%
\providecommand \@href[1]{\@@startlink{#1}\@@href}%
\providecommand \@@href[1]{\endgroup#1\@@endlink}%
\providecommand \@sanitize@url [0]{\catcode `\\12\catcode `\$12\catcode `\&12\catcode `\#12\catcode `\^12\catcode `\_12\catcode `\%12\relax}%
\providecommand \@@startlink[1]{}%
\providecommand \@@endlink[0]{}%
\providecommand \url  [0]{\begingroup\@sanitize@url \@url }%
\providecommand \@url [1]{\endgroup\@href {#1}{\urlprefix }}%
\providecommand \urlprefix  [0]{URL }%
\providecommand \Eprint [0]{\href }%
\providecommand \doibase [0]{https://doi.org/}%
\providecommand \selectlanguage [0]{\@gobble}%
\providecommand \bibinfo  [0]{\@secondoftwo}%
\providecommand \bibfield  [0]{\@secondoftwo}%
\providecommand \translation [1]{[#1]}%
\providecommand \BibitemOpen [0]{}%
\providecommand \bibitemStop [0]{}%
\providecommand \bibitemNoStop [0]{.\EOS\space}%
\providecommand \EOS [0]{\spacefactor3000\relax}%
\providecommand \BibitemShut  [1]{\csname bibitem#1\endcsname}%
\let\auto@bib@innerbib\@empty
\bibitem [{\citenamefont {McArdle}\ \emph {et~al.}(2020)\citenamefont {McArdle}, \citenamefont {Endo}, \citenamefont {Aspuru-Guzik}, \citenamefont {Benjamin},\ and\ \citenamefont {Yuan}}]{McArdle19}%
  \BibitemOpen
  \bibfield  {author} {\bibinfo {author} {\bibfnamefont {S.}~\bibnamefont {McArdle}}, \bibinfo {author} {\bibfnamefont {S.}~\bibnamefont {Endo}}, \bibinfo {author} {\bibfnamefont {A.}~\bibnamefont {Aspuru-Guzik}}, \bibinfo {author} {\bibfnamefont {S.~C.}\ \bibnamefont {Benjamin}},\ and\ \bibinfo {author} {\bibfnamefont {X.}~\bibnamefont {Yuan}},\ }\href {https://doi.org/10.1103/RevModPhys.92.015003} {\bibfield  {journal} {\bibinfo  {journal} {Rev. Mod. Phys.}\ }\textbf {\bibinfo {volume} {92}},\ \bibinfo {pages} {015003} (\bibinfo {year} {2020})}\BibitemShut {NoStop}%
\bibitem [{\citenamefont {Cao}\ \emph {et~al.}(2019)\citenamefont {Cao}, \citenamefont {Romero}, \citenamefont {Olson}, \citenamefont {Degroote}, \citenamefont {Johnson}, \citenamefont {Kieferová}, \citenamefont {Kivlichan}, \citenamefont {Menke}, \citenamefont {Peropadre}, \citenamefont {Sawaya}, \citenamefont {Sim}, \citenamefont {Veis},\ and\ \citenamefont {Aspuru-Guzik}}]{Cao19}%
  \BibitemOpen
  \bibfield  {author} {\bibinfo {author} {\bibfnamefont {Y.}~\bibnamefont {Cao}}, \bibinfo {author} {\bibfnamefont {J.}~\bibnamefont {Romero}}, \bibinfo {author} {\bibfnamefont {J.~P.}\ \bibnamefont {Olson}}, \bibinfo {author} {\bibfnamefont {M.}~\bibnamefont {Degroote}}, \bibinfo {author} {\bibfnamefont {P.~D.}\ \bibnamefont {Johnson}}, \bibinfo {author} {\bibfnamefont {M.}~\bibnamefont {Kieferová}}, \bibinfo {author} {\bibfnamefont {I.~D.}\ \bibnamefont {Kivlichan}}, \bibinfo {author} {\bibfnamefont {T.}~\bibnamefont {Menke}}, \bibinfo {author} {\bibfnamefont {B.}~\bibnamefont {Peropadre}}, \bibinfo {author} {\bibfnamefont {N.~P.~D.}\ \bibnamefont {Sawaya}}, \bibinfo {author} {\bibfnamefont {S.}~\bibnamefont {Sim}}, \bibinfo {author} {\bibfnamefont {L.}~\bibnamefont {Veis}},\ and\ \bibinfo {author} {\bibfnamefont {A.}~\bibnamefont {Aspuru-Guzik}},\ }\href {https://doi.org/10.1021/acs.chemrev.8b00803} {\bibfield  {journal} {\bibinfo  {journal} {Chem. Rev.}\ }\textbf {\bibinfo {volume} {119}},\ \bibinfo
  {pages} {10856} (\bibinfo {year} {2019})}\BibitemShut {NoStop}%
\bibitem [{\citenamefont {Auerbach}(1994)}]{Auerbach94}%
  \BibitemOpen
  \bibfield  {author} {\bibinfo {author} {\bibfnamefont {A.}~\bibnamefont {Auerbach}},\ }\href@noop {} {\emph {\bibinfo {title} {Interacting Electrons and Quantum Magnetism}}}\ (\bibinfo  {publisher} {Springer},\ \bibinfo {year} {1994})\BibitemShut {NoStop}%
\bibitem [{\citenamefont {Foulkes}\ \emph {et~al.}(2001)\citenamefont {Foulkes}, \citenamefont {Mitas}, \citenamefont {Needs},\ and\ \citenamefont {Rajagopal}}]{Foulkes01}%
  \BibitemOpen
  \bibfield  {author} {\bibinfo {author} {\bibfnamefont {W.~M.~C.}\ \bibnamefont {Foulkes}}, \bibinfo {author} {\bibfnamefont {L.}~\bibnamefont {Mitas}}, \bibinfo {author} {\bibfnamefont {R.~J.}\ \bibnamefont {Needs}},\ and\ \bibinfo {author} {\bibfnamefont {G.}~\bibnamefont {Rajagopal}},\ }\href {https://doi.org/10.1103/RevModPhys.73.33} {\bibfield  {journal} {\bibinfo  {journal} {Rev. Mod. Phys.}\ }\textbf {\bibinfo {volume} {73}},\ \bibinfo {pages} {33} (\bibinfo {year} {2001})}\BibitemShut {NoStop}%
\bibitem [{\citenamefont {Schollw\"ock}(2005)}]{Schollwock05}%
  \BibitemOpen
  \bibfield  {author} {\bibinfo {author} {\bibfnamefont {U.}~\bibnamefont {Schollw\"ock}},\ }\href {https://doi.org/10.1103/RevModPhys.77.259} {\bibfield  {journal} {\bibinfo  {journal} {Rev. Mod. Phys.}\ }\textbf {\bibinfo {volume} {77}},\ \bibinfo {pages} {259} (\bibinfo {year} {2005})}\BibitemShut {NoStop}%
\bibitem [{\citenamefont {Cade}\ \emph {et~al.}(2020)\citenamefont {Cade}, \citenamefont {Mineh}, \citenamefont {Montanaro},\ and\ \citenamefont {Stanisic}}]{Cade20}%
  \BibitemOpen
  \bibfield  {author} {\bibinfo {author} {\bibfnamefont {C.}~\bibnamefont {Cade}}, \bibinfo {author} {\bibfnamefont {L.}~\bibnamefont {Mineh}}, \bibinfo {author} {\bibfnamefont {A.}~\bibnamefont {Montanaro}},\ and\ \bibinfo {author} {\bibfnamefont {S.}~\bibnamefont {Stanisic}},\ }\href {https://doi.org/10.1103/PhysRevB.102.235122} {\bibfield  {journal} {\bibinfo  {journal} {Phys. Rev. B}\ }\textbf {\bibinfo {volume} {102}},\ \bibinfo {pages} {235122} (\bibinfo {year} {2020})}\BibitemShut {NoStop}%
\bibitem [{\citenamefont {Cai}(2020)}]{Cai20}%
  \BibitemOpen
  \bibfield  {author} {\bibinfo {author} {\bibfnamefont {Z.}~\bibnamefont {Cai}},\ }\href {https://doi.org/10.1103/PhysRevApplied.14.014059} {\bibfield  {journal} {\bibinfo  {journal} {Phys. Rev. Appl.}\ }\textbf {\bibinfo {volume} {14}},\ \bibinfo {pages} {014059} (\bibinfo {year} {2020})}\BibitemShut {NoStop}%
\bibitem [{\citenamefont {Preskill}(2018)}]{Preskill18}%
  \BibitemOpen
  \bibfield  {author} {\bibinfo {author} {\bibfnamefont {J.}~\bibnamefont {Preskill}},\ }\href {https://doi.org/10.22331/q-2018-08-06-79} {\bibfield  {journal} {\bibinfo  {journal} {{Quantum}}\ }\textbf {\bibinfo {volume} {2}},\ \bibinfo {pages} {79} (\bibinfo {year} {2018})}\BibitemShut {NoStop}%
\bibitem [{\citenamefont {Gutzwiller}(1963)}]{Gutzwiller63}%
  \BibitemOpen
  \bibfield  {author} {\bibinfo {author} {\bibfnamefont {M.~C.}\ \bibnamefont {Gutzwiller}},\ }\href {https://doi.org/10.1103/PhysRevLett.10.159} {\bibfield  {journal} {\bibinfo  {journal} {Phys. Rev. Lett.}\ }\textbf {\bibinfo {volume} {10}},\ \bibinfo {pages} {159} (\bibinfo {year} {1963})}\BibitemShut {NoStop}%
\bibitem [{\citenamefont {Hubbard}\ and\ \citenamefont {Flowers}(1963)}]{Hubbard63}%
  \BibitemOpen
  \bibfield  {author} {\bibinfo {author} {\bibfnamefont {J.}~\bibnamefont {Hubbard}}\ and\ \bibinfo {author} {\bibfnamefont {B.~H.}\ \bibnamefont {Flowers}},\ }\href {https://doi.org/10.1098/rspa.1963.0204} {\bibfield  {journal} {\bibinfo  {journal} {Proc. R. Soc. A: Math. Phys. Eng. Sci.}\ }\textbf {\bibinfo {volume} {276}},\ \bibinfo {pages} {238} (\bibinfo {year} {1963})}\BibitemShut {NoStop}%
\bibitem [{\citenamefont {Kanamori}(1963)}]{Kanamori63}%
  \BibitemOpen
  \bibfield  {author} {\bibinfo {author} {\bibfnamefont {J.}~\bibnamefont {Kanamori}},\ }\href {https://doi.org/10.1143/PTP.30.275} {\bibfield  {journal} {\bibinfo  {journal} {Prog. Theor. Phys.}\ }\textbf {\bibinfo {volume} {30}},\ \bibinfo {pages} {275} (\bibinfo {year} {1963})}\BibitemShut {NoStop}%
\bibitem [{\citenamefont {Arovas}\ \emph {et~al.}(2022)\citenamefont {Arovas}, \citenamefont {Berg}, \citenamefont {Kivelson},\ and\ \citenamefont {Raghu}}]{Arovas22}%
  \BibitemOpen
  \bibfield  {author} {\bibinfo {author} {\bibfnamefont {D.~P.}\ \bibnamefont {Arovas}}, \bibinfo {author} {\bibfnamefont {E.}~\bibnamefont {Berg}}, \bibinfo {author} {\bibfnamefont {S.~A.}\ \bibnamefont {Kivelson}},\ and\ \bibinfo {author} {\bibfnamefont {S.}~\bibnamefont {Raghu}},\ }\href {https://doi.org/10.1146/annurev-conmatphys-031620-102024} {\bibfield  {journal} {\bibinfo  {journal} {Annu. Rev. Condens. Matter Phys.}\ }\textbf {\bibinfo {volume} {13}},\ \bibinfo {pages} {239} (\bibinfo {year} {2022})}\BibitemShut {NoStop}%
\bibitem [{\citenamefont {Murta}\ and\ \citenamefont {Fern\'{a}ndez-Rossier}(2021)}]{Murta21}%
  \BibitemOpen
  \bibfield  {author} {\bibinfo {author} {\bibfnamefont {B.}~\bibnamefont {Murta}}\ and\ \bibinfo {author} {\bibfnamefont {J.}~\bibnamefont {Fern\'{a}ndez-Rossier}},\ }\href {https://doi.org/10.1103/PhysRevB.103.L241113} {\bibfield  {journal} {\bibinfo  {journal} {Phys. Rev. B}\ }\textbf {\bibinfo {volume} {103}},\ \bibinfo {pages} {L241113} (\bibinfo {year} {2021})}\BibitemShut {NoStop}%
\bibitem [{\citenamefont {Wecker}\ \emph {et~al.}(2015{\natexlab{a}})\citenamefont {Wecker}, \citenamefont {Hastings}, \citenamefont {Wiebe}, \citenamefont {Clark}, \citenamefont {Nayak},\ and\ \citenamefont {Troyer}}]{Wecker15}%
  \BibitemOpen
  \bibfield  {author} {\bibinfo {author} {\bibfnamefont {D.}~\bibnamefont {Wecker}}, \bibinfo {author} {\bibfnamefont {M.~B.}\ \bibnamefont {Hastings}}, \bibinfo {author} {\bibfnamefont {N.}~\bibnamefont {Wiebe}}, \bibinfo {author} {\bibfnamefont {B.~K.}\ \bibnamefont {Clark}}, \bibinfo {author} {\bibfnamefont {C.}~\bibnamefont {Nayak}},\ and\ \bibinfo {author} {\bibfnamefont {M.}~\bibnamefont {Troyer}},\ }\href {https://doi.org/10.1103/PhysRevA.92.062318} {\bibfield  {journal} {\bibinfo  {journal} {Phys. Rev. A}\ }\textbf {\bibinfo {volume} {92}},\ \bibinfo {pages} {062318} (\bibinfo {year} {2015}{\natexlab{a}})}\BibitemShut {NoStop}%
\bibitem [{\citenamefont {Seki}\ \emph {et~al.}(2022)\citenamefont {Seki}, \citenamefont {Otsuka},\ and\ \citenamefont {Yunoki}}]{Seki22}%
  \BibitemOpen
  \bibfield  {author} {\bibinfo {author} {\bibfnamefont {K.}~\bibnamefont {Seki}}, \bibinfo {author} {\bibfnamefont {Y.}~\bibnamefont {Otsuka}},\ and\ \bibinfo {author} {\bibfnamefont {S.}~\bibnamefont {Yunoki}},\ }\href {https://doi.org/10.1103/PhysRevB.105.155119} {\bibfield  {journal} {\bibinfo  {journal} {Phys. Rev. B}\ }\textbf {\bibinfo {volume} {105}},\ \bibinfo {pages} {155119} (\bibinfo {year} {2022})}\BibitemShut {NoStop}%
\bibitem [{\citenamefont {Heisenberg}(1928)}]{Heisenberg28}%
  \BibitemOpen
  \bibfield  {author} {\bibinfo {author} {\bibfnamefont {W.}~\bibnamefont {Heisenberg}},\ }\href {https://doi.org/10.1007/BF01328601} {\bibfield  {journal} {\bibinfo  {journal} {Z. Physik}\ }\textbf {\bibinfo {volume} {49}},\ \bibinfo {pages} {619} (\bibinfo {year} {1928})}\BibitemShut {NoStop}%
\bibitem [{\citenamefont {Kohn}(1999)}]{Kohn99}%
  \BibitemOpen
  \bibfield  {author} {\bibinfo {author} {\bibfnamefont {W.}~\bibnamefont {Kohn}},\ }\href {https://doi.org/10.1103/RevModPhys.71.1253} {\bibfield  {journal} {\bibinfo  {journal} {Rev. Mod. Phys.}\ }\textbf {\bibinfo {volume} {71}},\ \bibinfo {pages} {1253} (\bibinfo {year} {1999})}\BibitemShut {NoStop}%
\bibitem [{\citenamefont {Anderson}(1967)}]{Anderson67}%
  \BibitemOpen
  \bibfield  {author} {\bibinfo {author} {\bibfnamefont {P.~W.}\ \bibnamefont {Anderson}},\ }\href {https://doi.org/10.1103/PhysRevLett.18.1049} {\bibfield  {journal} {\bibinfo  {journal} {Phys. Rev. Lett.}\ }\textbf {\bibinfo {volume} {18}},\ \bibinfo {pages} {1049} (\bibinfo {year} {1967})}\BibitemShut {NoStop}%
\bibitem [{\citenamefont {McClean}\ \emph {et~al.}(2018)\citenamefont {McClean}, \citenamefont {Boixo}, \citenamefont {Smelyanskiy}, \citenamefont {Babbush},\ and\ \citenamefont {Neven}}]{McClean18}%
  \BibitemOpen
  \bibfield  {author} {\bibinfo {author} {\bibfnamefont {J.~R.}\ \bibnamefont {McClean}}, \bibinfo {author} {\bibfnamefont {S.}~\bibnamefont {Boixo}}, \bibinfo {author} {\bibfnamefont {V.~N.}\ \bibnamefont {Smelyanskiy}}, \bibinfo {author} {\bibfnamefont {R.}~\bibnamefont {Babbush}},\ and\ \bibinfo {author} {\bibfnamefont {H.}~\bibnamefont {Neven}},\ }\href {http://dx.doi.org/10.1038/s41467-018-07090-4} {\bibfield  {journal} {\bibinfo  {journal} {Nat. Commun.}\ }\textbf {\bibinfo {volume} {9}},\ \bibinfo {pages} {1} (\bibinfo {year} {2018})}\BibitemShut {NoStop}%
\bibitem [{\citenamefont {Cerezo}\ \emph {et~al.}(2021)\citenamefont {Cerezo}, \citenamefont {Arrasmith}, \citenamefont {Babbush}, \citenamefont {Benjamin}, \citenamefont {Endo}, \citenamefont {Fujii}, \citenamefont {McClean}, \citenamefont {Mitarai}, \citenamefont {Yuan}, \citenamefont {Cincio},\ and\ \citenamefont {Coles}}]{Cerezo21}%
  \BibitemOpen
  \bibfield  {author} {\bibinfo {author} {\bibfnamefont {M.}~\bibnamefont {Cerezo}}, \bibinfo {author} {\bibfnamefont {A.}~\bibnamefont {Arrasmith}}, \bibinfo {author} {\bibfnamefont {R.}~\bibnamefont {Babbush}}, \bibinfo {author} {\bibfnamefont {S.~C.}\ \bibnamefont {Benjamin}}, \bibinfo {author} {\bibfnamefont {S.}~\bibnamefont {Endo}}, \bibinfo {author} {\bibfnamefont {K.}~\bibnamefont {Fujii}}, \bibinfo {author} {\bibfnamefont {J.~R.}\ \bibnamefont {McClean}}, \bibinfo {author} {\bibfnamefont {K.}~\bibnamefont {Mitarai}}, \bibinfo {author} {\bibfnamefont {X.}~\bibnamefont {Yuan}}, \bibinfo {author} {\bibfnamefont {L.}~\bibnamefont {Cincio}},\ and\ \bibinfo {author} {\bibfnamefont {P.~J.}\ \bibnamefont {Coles}},\ }\href {https://doi.org/10.1038/s42254-021-00348-9} {\bibfield  {journal} {\bibinfo  {journal} {Nat. Rev. Phys.}\ }\textbf {\bibinfo {volume} {3}},\ \bibinfo {pages} {625} (\bibinfo {year} {2021})}\BibitemShut {NoStop}%
\bibitem [{\citenamefont {Peruzzo}\ \emph {et~al.}(2014)\citenamefont {Peruzzo}, \citenamefont {McClean}, \citenamefont {Shadbolt}, \citenamefont {Yung}, \citenamefont {Zhou}, \citenamefont {Love}, \citenamefont {Aspuru-Guzik},\ and\ \citenamefont {O'Brien}}]{Peruzzo14}%
  \BibitemOpen
  \bibfield  {author} {\bibinfo {author} {\bibfnamefont {A.}~\bibnamefont {Peruzzo}}, \bibinfo {author} {\bibfnamefont {J.}~\bibnamefont {McClean}}, \bibinfo {author} {\bibfnamefont {P.}~\bibnamefont {Shadbolt}}, \bibinfo {author} {\bibfnamefont {M.-H.}\ \bibnamefont {Yung}}, \bibinfo {author} {\bibfnamefont {X.-Q.}\ \bibnamefont {Zhou}}, \bibinfo {author} {\bibfnamefont {P.~J.}\ \bibnamefont {Love}}, \bibinfo {author} {\bibfnamefont {A.}~\bibnamefont {Aspuru-Guzik}},\ and\ \bibinfo {author} {\bibfnamefont {J.~L.}\ \bibnamefont {O'Brien}},\ }\href {https://doi.org/10.1038/ncomms5213} {\bibfield  {journal} {\bibinfo  {journal} {Nat. Commun.}\ }\textbf {\bibinfo {volume} {5}},\ \bibinfo {pages} {4213} (\bibinfo {year} {2014})}\BibitemShut {NoStop}%
\bibitem [{\citenamefont {Seki}\ \emph {et~al.}(2020)\citenamefont {Seki}, \citenamefont {Shirakawa},\ and\ \citenamefont {Yunoki}}]{Seki20}%
  \BibitemOpen
  \bibfield  {author} {\bibinfo {author} {\bibfnamefont {K.}~\bibnamefont {Seki}}, \bibinfo {author} {\bibfnamefont {T.}~\bibnamefont {Shirakawa}},\ and\ \bibinfo {author} {\bibfnamefont {S.}~\bibnamefont {Yunoki}},\ }\href {https://doi.org/10.1103/PhysRevA.101.052340} {\bibfield  {journal} {\bibinfo  {journal} {Phys. Rev. A}\ }\textbf {\bibinfo {volume} {101}},\ \bibinfo {pages} {052340} (\bibinfo {year} {2020})}\BibitemShut {NoStop}%
\bibitem [{\citenamefont {Jordan}\ and\ \citenamefont {Wigner}(1928)}]{JordanWigner28}%
  \BibitemOpen
  \bibfield  {author} {\bibinfo {author} {\bibfnamefont {P.}~\bibnamefont {Jordan}}\ and\ \bibinfo {author} {\bibfnamefont {E.}~\bibnamefont {Wigner}},\ }\href {https://doi.org/10.1007/BF01331938} {\bibfield  {journal} {\bibinfo  {journal} {Z. Phys.}\ }\textbf {\bibinfo {volume} {47}},\ \bibinfo {pages} {631} (\bibinfo {year} {1928})}\BibitemShut {NoStop}%
\bibitem [{\citenamefont {Lieb}\ and\ \citenamefont {Mattis}(1962)}]{LiebMattis62}%
  \BibitemOpen
  \bibfield  {author} {\bibinfo {author} {\bibfnamefont {E.}~\bibnamefont {Lieb}}\ and\ \bibinfo {author} {\bibfnamefont {D.}~\bibnamefont {Mattis}},\ }\href {https://doi.org/10.1063/1.1724276} {\bibfield  {journal} {\bibinfo  {journal} {J. Math. Phys.}\ }\textbf {\bibinfo {volume} {3}},\ \bibinfo {pages} {749} (\bibinfo {year} {1962})}\BibitemShut {NoStop}%
\bibitem [{\citenamefont {Bosse}\ and\ \citenamefont {Montanaro}(2022)}]{BosseMontanaro22}%
  \BibitemOpen
  \bibfield  {author} {\bibinfo {author} {\bibfnamefont {J.~L.}\ \bibnamefont {Bosse}}\ and\ \bibinfo {author} {\bibfnamefont {A.}~\bibnamefont {Montanaro}},\ }\href {https://doi.org/10.1103/PhysRevB.105.094409} {\bibfield  {journal} {\bibinfo  {journal} {Phys. Rev. B}\ }\textbf {\bibinfo {volume} {105}},\ \bibinfo {pages} {094409} (\bibinfo {year} {2022})}\BibitemShut {NoStop}%
\bibitem [{\citenamefont {Yu}\ \emph {et~al.}(2023)\citenamefont {Yu}, \citenamefont {Zhao},\ and\ \citenamefont {Wei}}]{YuZhaoWei23}%
  \BibitemOpen
  \bibfield  {author} {\bibinfo {author} {\bibfnamefont {H.}~\bibnamefont {Yu}}, \bibinfo {author} {\bibfnamefont {Y.}~\bibnamefont {Zhao}},\ and\ \bibinfo {author} {\bibfnamefont {T.-C.}\ \bibnamefont {Wei}},\ }\href {https://doi.org/10.1103/PhysRevResearch.5.013183} {\bibfield  {journal} {\bibinfo  {journal} {Phys. Rev. Res.}\ }\textbf {\bibinfo {volume} {5}},\ \bibinfo {pages} {013183} (\bibinfo {year} {2023})}\BibitemShut {NoStop}%
\bibitem [{\citenamefont {Farhi}\ \emph {et~al.}(2014)\citenamefont {Farhi}, \citenamefont {Goldstone},\ and\ \citenamefont {Gutmann}}]{Farhi14}%
  \BibitemOpen
  \bibfield  {author} {\bibinfo {author} {\bibfnamefont {E.}~\bibnamefont {Farhi}}, \bibinfo {author} {\bibfnamefont {J.}~\bibnamefont {Goldstone}},\ and\ \bibinfo {author} {\bibfnamefont {S.}~\bibnamefont {Gutmann}},\ }\href {https://arxiv.org/abs/1411.4028} {\bibfield  {journal} {\bibinfo  {journal} {arXiv:1411.4028}\ } (\bibinfo {year} {2014})}\BibitemShut {NoStop}%
\bibitem [{\citenamefont {Wecker}\ \emph {et~al.}(2015{\natexlab{b}})\citenamefont {Wecker}, \citenamefont {Hastings},\ and\ \citenamefont {Troyer}}]{Wecker15a}%
  \BibitemOpen
  \bibfield  {author} {\bibinfo {author} {\bibfnamefont {D.}~\bibnamefont {Wecker}}, \bibinfo {author} {\bibfnamefont {M.~B.}\ \bibnamefont {Hastings}},\ and\ \bibinfo {author} {\bibfnamefont {M.}~\bibnamefont {Troyer}},\ }\href {https://doi.org/10.1103/PhysRevA.92.042303} {\bibfield  {journal} {\bibinfo  {journal} {Phys. Rev. A}\ }\textbf {\bibinfo {volume} {92}},\ \bibinfo {pages} {042303} (\bibinfo {year} {2015}{\natexlab{b}})}\BibitemShut {NoStop}%
\bibitem [{\citenamefont {Anderson}(1973)}]{Anderson73}%
  \BibitemOpen
  \bibfield  {author} {\bibinfo {author} {\bibfnamefont {P.}~\bibnamefont {Anderson}},\ }\href {https://doi.org/https://doi.org/10.1016/0025-5408(73)90167-0} {\bibfield  {journal} {\bibinfo  {journal} {Mater. Res. Bull.}\ }\textbf {\bibinfo {volume} {8}},\ \bibinfo {pages} {153} (\bibinfo {year} {1973})}\BibitemShut {NoStop}%
\bibitem [{\citenamefont {Anderson}(1987)}]{Anderson87}%
  \BibitemOpen
  \bibfield  {author} {\bibinfo {author} {\bibfnamefont {P.~W.}\ \bibnamefont {Anderson}},\ }\href {https://doi.org/10.1126/science.235.4793.1196} {\bibfield  {journal} {\bibinfo  {journal} {Science}\ }\textbf {\bibinfo {volume} {235}},\ \bibinfo {pages} {1196} (\bibinfo {year} {1987})}\BibitemShut {NoStop}%
\bibitem [{Sup()}]{SuppMatRef}%
  \BibitemOpen
  \href@noop {} {\bibinfo  {journal} {{See Supplemental Material at [Add URL] for the basis gate decomposition of all quantum circuits considered in this Letter, details about the optimization of the free parameters in the VQE simulations, additional information about the conversion of a multi-spin-$\frac{1}{2}$ wave function into its fermionic version, the illustration with the Fermi-Hubbard dimer of the need to have the Z strings in the spin-$\uparrow$ hopping terms of Eq. (\ref{eq:D_theta_fermion}) for the addition of doublon-hole pairs to produce an improvement of the overlap with the exact ground state, and analogous results to those presented in Fig. \ref{fig3} for smaller lattices with the same geometry}}\ }\BibitemShut {NoStop}%
\bibitem [{\citenamefont {Weinberg}\ and\ \citenamefont {Bukov}(2017)}]{QuSpin}%
  \BibitemOpen
\bibfield  {journal} {  }\bibfield  {author} {\bibinfo {author} {\bibfnamefont {P.}~\bibnamefont {Weinberg}}\ and\ \bibinfo {author} {\bibfnamefont {M.}~\bibnamefont {Bukov}},\ }\href {https://quspin.github.io/QuSpin/} {\bibinfo {title} {Qu{S}pin: a {P}ython package for dynamics and exact diagonalization of quantum many-body systems}} (\bibinfo {year} {2017}),\ \bibinfo {note} {last accessed on 16-10-2023}\BibitemShut {NoStop}%
\bibitem [{\citenamefont {Kivlichan}\ \emph {et~al.}(2018)\citenamefont {Kivlichan} \emph {et~al.}}]{Kivlichan18}%
  \BibitemOpen
  \bibfield  {author} {\bibinfo {author} {\bibfnamefont {I.~D.}\ \bibnamefont {Kivlichan}} \emph {et~al.},\ }\href {https://doi.org/10.1103/PhysRevLett.120.110501} {\bibfield  {journal} {\bibinfo  {journal} {Phys. Rev. Lett.}\ }\textbf {\bibinfo {volume} {120}},\ \bibinfo {pages} {110501} (\bibinfo {year} {2018})}\BibitemShut {NoStop}%
\bibitem [{\citenamefont {Jiang}\ \emph {et~al.}(2018)\citenamefont {Jiang}, \citenamefont {Sung}, \citenamefont {Kechedzhi}, \citenamefont {Smelyanskiy},\ and\ \citenamefont {Boixo}}]{Jiang18}%
  \BibitemOpen
  \bibfield  {author} {\bibinfo {author} {\bibfnamefont {Z.}~\bibnamefont {Jiang}}, \bibinfo {author} {\bibfnamefont {K.~J.}\ \bibnamefont {Sung}}, \bibinfo {author} {\bibfnamefont {K.}~\bibnamefont {Kechedzhi}}, \bibinfo {author} {\bibfnamefont {V.~N.}\ \bibnamefont {Smelyanskiy}},\ and\ \bibinfo {author} {\bibfnamefont {S.}~\bibnamefont {Boixo}},\ }\href {https://doi.org/10.1103/PhysRevApplied.9.044036} {\bibfield  {journal} {\bibinfo  {journal} {Phys. Rev. Appl.}\ }\textbf {\bibinfo {volume} {9}},\ \bibinfo {pages} {044036} (\bibinfo {year} {2018})}\BibitemShut {NoStop}%
\bibitem [{\citenamefont {Verstraete}\ \emph {et~al.}(2009)\citenamefont {Verstraete}, \citenamefont {Cirac},\ and\ \citenamefont {Latorre}}]{Verstraete09}%
  \BibitemOpen
  \bibfield  {author} {\bibinfo {author} {\bibfnamefont {F.}~\bibnamefont {Verstraete}}, \bibinfo {author} {\bibfnamefont {J.~I.}\ \bibnamefont {Cirac}},\ and\ \bibinfo {author} {\bibfnamefont {J.~I.}\ \bibnamefont {Latorre}},\ }\href {https://doi.org/10.1103/PhysRevA.79.032316} {\bibfield  {journal} {\bibinfo  {journal} {Phys. Rev. A}\ }\textbf {\bibinfo {volume} {79}},\ \bibinfo {pages} {032316} (\bibinfo {year} {2009})}\BibitemShut {NoStop}%
\bibitem [{\citenamefont {Baeriswyl}(2000)}]{Baeriswyl00}%
  \BibitemOpen
  \bibfield  {author} {\bibinfo {author} {\bibfnamefont {D.}~\bibnamefont {Baeriswyl}},\ }\href {https://doi.org/10.1023/A:1003785323041} {\bibfield  {journal} {\bibinfo  {journal} {Found. Phys.}\ }\textbf {\bibinfo {volume} {30}},\ \bibinfo {pages} {2033} (\bibinfo {year} {2000})}\BibitemShut {NoStop}%
\bibitem [{\citenamefont {Bethe}(1931)}]{Bethe31}%
  \BibitemOpen
  \bibfield  {author} {\bibinfo {author} {\bibfnamefont {H.}~\bibnamefont {Bethe}},\ }\href {https://doi.org/10.1007/BF01341708} {\bibfield  {journal} {\bibinfo  {journal} {Z. Phys.}\ }\textbf {\bibinfo {volume} {71}},\ \bibinfo {pages} {205} (\bibinfo {year} {1931})}\BibitemShut {NoStop}%
\bibitem [{\citenamefont {Van~Dyke}\ \emph {et~al.}(2021)\citenamefont {Van~Dyke}, \citenamefont {Barron}, \citenamefont {Mayhall}, \citenamefont {Barnes},\ and\ \citenamefont {Economou}}]{VanDyke21}%
  \BibitemOpen
  \bibfield  {author} {\bibinfo {author} {\bibfnamefont {J.~S.}\ \bibnamefont {Van~Dyke}}, \bibinfo {author} {\bibfnamefont {G.~S.}\ \bibnamefont {Barron}}, \bibinfo {author} {\bibfnamefont {N.~J.}\ \bibnamefont {Mayhall}}, \bibinfo {author} {\bibfnamefont {E.}~\bibnamefont {Barnes}},\ and\ \bibinfo {author} {\bibfnamefont {S.~E.}\ \bibnamefont {Economou}},\ }\href {https://doi.org/10.1103/PRXQuantum.2.040329} {\bibfield  {journal} {\bibinfo  {journal} {PRX Quantum}\ }\textbf {\bibinfo {volume} {2}},\ \bibinfo {pages} {040329} (\bibinfo {year} {2021})}\BibitemShut {NoStop}%
\bibitem [{\citenamefont {Sopena}\ \emph {et~al.}(2022)\citenamefont {Sopena}, \citenamefont {Gordon}, \citenamefont {Garc{\'{i}}a-Mart{\'{i}}n}, \citenamefont {Sierra},\ and\ \citenamefont {L{\'{o}}pez}}]{Sopena22}%
  \BibitemOpen
  \bibfield  {author} {\bibinfo {author} {\bibfnamefont {A.}~\bibnamefont {Sopena}}, \bibinfo {author} {\bibfnamefont {M.~H.}\ \bibnamefont {Gordon}}, \bibinfo {author} {\bibfnamefont {D.}~\bibnamefont {Garc{\'{i}}a-Mart{\'{i}}n}}, \bibinfo {author} {\bibfnamefont {G.}~\bibnamefont {Sierra}},\ and\ \bibinfo {author} {\bibfnamefont {E.}~\bibnamefont {L{\'{o}}pez}},\ }\href {https://doi.org/10.22331/q-2022-09-08-796} {\bibfield  {journal} {\bibinfo  {journal} {{Quantum}}\ }\textbf {\bibinfo {volume} {6}},\ \bibinfo {pages} {796} (\bibinfo {year} {2022})}\BibitemShut {NoStop}%
\bibitem [{\citenamefont {Majumdar}\ and\ \citenamefont {Ghosh}(1969{\natexlab{a}})}]{MajumdarGhosh69a}%
  \BibitemOpen
  \bibfield  {author} {\bibinfo {author} {\bibfnamefont {C.~K.}\ \bibnamefont {Majumdar}}\ and\ \bibinfo {author} {\bibfnamefont {D.~K.}\ \bibnamefont {Ghosh}},\ }\href {https://doi.org/10.1063/1.1664978} {\bibfield  {journal} {\bibinfo  {journal} {J. Math. Phys.}\ }\textbf {\bibinfo {volume} {10}},\ \bibinfo {pages} {1388} (\bibinfo {year} {1969}{\natexlab{a}})}\BibitemShut {NoStop}%
\bibitem [{\citenamefont {Majumdar}\ and\ \citenamefont {Ghosh}(1969{\natexlab{b}})}]{MajumdarGhosh69b}%
  \BibitemOpen
  \bibfield  {author} {\bibinfo {author} {\bibfnamefont {C.~K.}\ \bibnamefont {Majumdar}}\ and\ \bibinfo {author} {\bibfnamefont {D.~K.}\ \bibnamefont {Ghosh}},\ }\href {https://doi.org/10.1063/1.1664979} {\bibfield  {journal} {\bibinfo  {journal} {J. Math. Phys.}\ }\textbf {\bibinfo {volume} {10}},\ \bibinfo {pages} {1399} (\bibinfo {year} {1969}{\natexlab{b}})}\BibitemShut {NoStop}%
\bibitem [{\citenamefont {{Sriram Shastry}}\ and\ \citenamefont {Sutherland}(1981)}]{ShastrySutherland81}%
  \BibitemOpen
  \bibfield  {author} {\bibinfo {author} {\bibfnamefont {B.}~\bibnamefont {{Sriram Shastry}}}\ and\ \bibinfo {author} {\bibfnamefont {B.}~\bibnamefont {Sutherland}},\ }\href {https://doi.org/https://doi.org/10.1016/0378-4363(81)90838-X} {\bibfield  {journal} {\bibinfo  {journal} {Physica B+C}\ }\textbf {\bibinfo {volume} {108}},\ \bibinfo {pages} {1069} (\bibinfo {year} {1981})}\BibitemShut {NoStop}%
\bibitem [{\citenamefont {Corboz}\ and\ \citenamefont {Mila}(2013)}]{Corboz13}%
  \BibitemOpen
  \bibfield  {author} {\bibinfo {author} {\bibfnamefont {P.}~\bibnamefont {Corboz}}\ and\ \bibinfo {author} {\bibfnamefont {F.}~\bibnamefont {Mila}},\ }\href {https://doi.org/10.1103/PhysRevB.87.115144} {\bibfield  {journal} {\bibinfo  {journal} {Phys. Rev. B}\ }\textbf {\bibinfo {volume} {87}},\ \bibinfo {pages} {115144} (\bibinfo {year} {2013})}\BibitemShut {NoStop}%
\bibitem [{\citenamefont {Cross}\ \emph {et~al.}(2017)\citenamefont {Cross}, \citenamefont {Bishop}, \citenamefont {Smolin},\ and\ \citenamefont {Gambetta}}]{OpenQASM17}%
  \BibitemOpen
  \bibfield  {author} {\bibinfo {author} {\bibfnamefont {A.~W.}\ \bibnamefont {Cross}}, \bibinfo {author} {\bibfnamefont {L.~S.}\ \bibnamefont {Bishop}}, \bibinfo {author} {\bibfnamefont {J.~A.}\ \bibnamefont {Smolin}},\ and\ \bibinfo {author} {\bibfnamefont {J.~M.}\ \bibnamefont {Gambetta}},\ }\href {https://arxiv.org/abs/1707.03429} {\bibfield  {journal} {\bibinfo  {journal} {arXiv:1707.03429}\ } (\bibinfo {year} {2017})}\BibitemShut {NoStop}%
\bibitem [{\citenamefont {Sakurai}\ and\ \citenamefont {Napolitano}(2017)}]{SakuraiNapolitano17}%
  \BibitemOpen
  \bibfield  {author} {\bibinfo {author} {\bibfnamefont {J.~J.}\ \bibnamefont {Sakurai}}\ and\ \bibinfo {author} {\bibfnamefont {J.}~\bibnamefont {Napolitano}},\ }\href {https://doi.org/10.1017/9781108499996} {\emph {\bibinfo {title} {{Modern Quantum Mechanics}}}},\ \bibinfo {edition} {2nd}\ ed.\ (\bibinfo  {publisher} {Cambridge University Press},\ \bibinfo {year} {2017})\BibitemShut {NoStop}%
\bibitem [{\citenamefont {{The Scipy Community}}(2008)}]{scipyminimize}%
  \BibitemOpen
  \bibfield  {author} {\bibinfo {author} {\bibnamefont {{The Scipy Community}}},\ }\href {https://docs.scipy.org/doc/scipy/reference/generated/scipy.optimize.minimize.html} {\bibinfo {title} {{SciPy Optimize: minimize Documentation}}} (\bibinfo {year} {2008}),\ \bibinfo {note} {last accessed on 16-10-2023}\BibitemShut {NoStop}%
\bibitem [{\citenamefont {Cruz}\ and\ \citenamefont {Murta}(2023)}]{CruzMurta23}%
  \BibitemOpen
  \bibfield  {author} {\bibinfo {author} {\bibfnamefont {P.~M.~Q.}\ \bibnamefont {Cruz}}\ and\ \bibinfo {author} {\bibfnamefont {B.}~\bibnamefont {Murta}},\ }\href {https://arxiv.org/abs/2305.18128} {\bibfield  {journal} {\bibinfo  {journal} {arXiv:2305.18128}\ } (\bibinfo {year} {2023})}\BibitemShut {NoStop}%
\bibitem [{\citenamefont {Nielsen}\ and\ \citenamefont {Chuang}(2010)}]{Nielsen00}%
  \BibitemOpen
  \bibfield  {author} {\bibinfo {author} {\bibfnamefont {M.~A.}\ \bibnamefont {Nielsen}}\ and\ \bibinfo {author} {\bibfnamefont {I.~L.}\ \bibnamefont {Chuang}},\ }\href {https://doi.org/10.1017/CBO9780511976667} {\emph {\bibinfo {title} {{Quantum Computation and Quantum Information: 10th Anniversary Edition}}}}\ (\bibinfo  {publisher} {{Cambridge University Press}},\ \bibinfo {year} {2010})\BibitemShut {NoStop}%
\end{thebibliography}%

\newpage

\end{document}